\documentclass[aps,prd,onecolumn,groupedaddress,showpacs,nofootinbib,amssymb,amsmath, floats, floatfix]{revtex4}

\usepackage[latin1]{inputenc}
\usepackage{graphicx}
\usepackage[english]{babel}
\usepackage{amsmath,amssymb}
\usepackage{amsfonts}
\usepackage{xspace} 
\usepackage{dcolumn}
\usepackage{bm}

\providecommand{\boldsymbol}[1]{\mbox{\boldmath $#1$}}

\begin{document}

\title{Gravitomagnetic corrections on gravitational waves}

\author{ S. Capozziello, M. De Laurentis, L. Forte, F. Garufi, and L. Milano}

\affiliation{\it Dipartimento di Scienze Fisiche, Università di
Napoli {}`` Federico II'' and INFN Sez. di Napoli, Compl. Univ. di
Monte S. Angelo, Edificio G, Via Cinthia, I-80126, Napoli, Italy}
\date{\today}

\begin{abstract}
Gravitational waveforms and  production  could be considerably
affected by gravitomagnetic corrections  considered in
relativistic theory of orbits. Beside the standard periastron
effect of General Relativity,  new nutation effects come out when
${\displaystyle c^{-3}}$ corrections are taken into account. Such
corrections emerge as soon as matter-current densities and vector
gravitational potentials cannot be discarded into dynamics. We
study the gravitational waves emitted through the capture, in the
gravitational field of massive binary systems (e.g. a very massive
black hole on which a stellar object is inspiralling) via the
quadrupole approximation, considering precession and nutation
effects. We present a numerical study to obtain the gravitational
wave luminosity, the total energy output and the gravitational
radiation amplitude. From a crude estimate of the expected number
of events towards peculiar targets (e.g. globular clusters) and in
particular, the rate of events per year for dense stellar clusters
at the Galactic Center (SgrA$^*$), we conclude that this type of
capture could give  signatures to be revealed by interferometric
GW antennas, in particular by the forthcoming laser interferometer
space antenna LISA.
\end{abstract}


\maketitle

%
%

\section{\label{intro} Introduction }
Searching for signatures of gravitational waves (GWs) and
achieving a suitable classification of  emitting sources have
become two crucial tasks in GW-science. In fact, the today
sensitivity levels and theoretical developments are leading toward
a general picture of GW-phenomena which could not be possible in
the previous pioneering era. Experimentally, several GW
ground-based-laser-interferometer detectors ($10^{-1}kHz$) have
been built in the United States (LIGO) \cite{Abra}, Europe (VIRGO
and GEO) \cite{Caron,Luck} and Japan (TAMA) \cite{Ando}, and are
now taking data at designed sensitivities. A laser-interferometer
space antenna (LISA) \cite{LISA} ($10^{-4}\sim10^{-2} Hz$) might
fly within the next decade.

From a theoretical point of view, recent years have been
characterized by numerous major advances due, essentially, to the
development of numerical gravity. Concerning the most promising
sources to be detected, the GW generation problem has improved
significantly in relation to the dynamics of binary and multiple
systems of compact objects as neutron stars and black holes.
Besides, the problem of non-geodesic motion of particles in curved
spacetime has been developed considering  the emission of GWs
\cite{Poisson,Mino}. Solving these problems is of considerable
importance in order to predict  the accurate waveforms of GWs
emitted by extreme mass-ratio binaries, which are among the most
promising sources for LISA \cite{Finn}.

From a more genuine astrophysical viewpoint, observations towards
the central regions of galaxies have detected  peculiar compact
massive objects which are present in almost all observed galaxies.
The occurrence of such systems has been revealed thanks to the
advance in high angular resolution instrumentation for a wide
range of electromagnetic wavelengths. Space-telescopes as HST or
ground-based telescope, which use adaptive optics, have been
extremely useful for studying kinematics of galactic internal
regions reaching accuracy of mill-pc for the Milky Way and of
pc-fractions for external galaxies. The main conclusion of all
these studies is that the central region of most of galaxies is
dominated by large compact objects with masses of the order
$M\simeq 10^6\div10^9 M_{\odot}$. In the case of Milky Way, the
peculiar object in the Sagittarius region (SgrA$^*$) is of the
order $M\simeq 3\times 10^6 M_{\odot}$ and it is usually addressed
as a massive black hole (MBH), also if its true physical nature is
far to be finally identified \cite{viollier,pla}.

In any case, a deep link exists between the central MBH and the
geometrical, kinematical and dynamical features of the host
galaxy. In particular, the MBH is correlated with the global shape
of galactic spheroid, with the velocity dispersion of surrounding
stars, with the mean density and the total mass of the host
galaxy. Dynamics of stars moving around the MBH has a series of
interesting characteristics which are of extreme interest for GW
detection and production. Due to this occurrence, searching for
GWs coming from objects interacting with MBHs is a major task for
GW interferometry from space and ground-based experiments.

In this paper, we are going to study the evolution of  compact
binary systems,  formed through the  capture of a moving (stellar)
mass $m$ by the gravitational field, whose source is a massive MBH
of mass $M$ where $m \ll M$.  One expects that small compact
objects ($1\div 20 M_{\odot}$) from the surrounding stellar
population will be captured by these black holes following
many-body scattering interactions at a relatively high rate
\cite{Sigurdsson,Sigurdsson2}. It is well known that the  capture
of stellar-mass compact objects by massive  MBHs could constitute,
potentially, a very important target for LISA
\cite{Danzmann,freitag}. However, dynamics has to be carefully
discussed in order to consider and select all effects coming from
standard stellar mass objects inspiralling  over MBHs.

In a previous paper \cite{Capoz}, we have shown that, in the
relativistic weak field approximation, when considering higher
order corrections to the equations of motion, gravitomagnetic
effects in the theory of orbits,  can be particularly significant,
leading also to chaotic behaviors in the transient regime dividing
stable from unstable trajectories. Generally, such contributions
are discarded since they are considered too small. However, in a
more accurate analysis, this is not true and gravitomagnetic
corrections could give peculiar characterization of dynamics.

In  \cite{Capoz},  Newtonian and  relativistic theories of orbits
have been reviewed  considering, in particular, how relativistic
corrections affect the "classical" orbits \cite{binney,landau}.
Equations of motion and phase portraits of solutions indicate
that, beside the standard periastron precession at order $c^{-2}$,
new nutation effects come out at order $c^{-3}$ and it is
misleading to neglect them.

According to these effects, orbits remain rather eccentric  until
the final plunge, and display both extreme  relativistic
perihelion precession and Lense-Thirring
\cite{Thirring1,Thirring,iorio} precession of the orbital plane
due to the spin of MBH, as well as orbital decay. In \cite{Ryan},
it is illustrated how the measured GW-waveforms can  effectively
map out the spacetime geometry close to the MBH. In
\cite{DeLaurentis,nucita},   the classical orbital motion (without
relativistic corrections in the motion of the binary system) has
been studied  in the extreme mass ratio limit $m\ll M$, assuming
the stellar system density and richness as fundamental parameters.
The conclusions have been  that $i)$ the GW-waveforms have
characterized by the orbital motion (in particular, closed or open
orbits give rise to very different GW-production and waveform
shapes); $ii)$  in rich and dense stellar clusters, a large
production of GWs can be expected, so that these systems could be
very interesting for the above mentioned ground-based and space
detectors; $iii)$ the amplitudes of the strongest GW signals are
expected to be roughly an order of magnitude smaller than LISA's
instrumental noise.

In this  paper, we investigate the GW emission by binary systems,
in the extreme  mass ratio limit, by the quadrupole approximation,
considering  orbits affected by both nutation and precession
effects,  taking into account also  gravitomagnetic terms in the
weak field approximation of the metric. We will see that
gravitational waves are emitted with a "peculiar" signature
related to the orbital features: such a signature may be a "burst"
wave-form with a maximum in correspondence to the periastron
distance or a modulated waveform, according to the orbit
stability. Here we face this problem discussing in detail the
dynamics of such a phenomenon which could greatly improve the
statistics of possible GW sources.

Besides, we give estimates of the distributions of these sources
and their parameters. It is worth noticing  that the captures
occur when  objects, in the dense stellar cusp surrounding a
galactic MBH, undergo a close encounter, so that the trajectory
becomes tight enough that orbital decay through emission of GWs
dominates the subsequent evolution. According to Refs.
\cite{Cutler1,Cutler2}), for a typical capture, the initial
orbital eccentricity is extremely large (typically $1-e\sim
10^{-6}{-}10^{-3}$) and the initial pericenter distance very small
($r_{\rm p}\sim 8-100 M$, where $M$ is the MBH mass
\cite{FreitagApJ}. The subsequent orbital evolution may (very
roughly) be divided into three stages. In the first and longest
stage the orbit is extremely eccentric, and GWs are emitted in
short ``pulses'' during pericenter passages. These GW pulses
slowly remove energy and angular momentum from the system, and the
orbit gradually shrinks and circularizes. After $\sim 10^3-10^8$
years (depending on the two masses and the initial eccentricity)
the evolution enters its second stage, where the orbit is
sufficiently circular:  the emission can be viewed as continuous.
Finally, as the object reaches the last stable orbit, the
adiabatic inspiral transits to a direct plunge, and the GW signal
cuts off. Radiation reaction quickly circularizes the orbit over
the inspiral phase; however, initial eccentricities are large
enough that a substantial fraction of captures will maintain high
eccentricity  until the final plunge. It has been estimated
\cite{Cutler1} that about half of the captures will plunge with
eccentricity $e\gtrsim 0.2$. While individually-resolvable
captures will mostly be detectable during the last $\sim 1-100$
yrs of the second stage (depending on the stellar mass $m$ and the
MBH mass), radiation emitted during the first stage will
contribute significantly to the confusion background. As we shall
see,  the above scenario is heavily modified since the
gravitomagnetic effects play a crucial role in modify the orbital
shapes that are far from being simply circular or elliptic and no
longer closed.

The layout of the paper is the following. In Sect.II, we give a
summary of gravitomagnetic corrections to the metric showing how
the geodesic equation results modified by their presence. Besides
we study in details orbits with such corrections showing the phase
portraits and the velocity fields determined by the motion of mass
$m$ around the MBH.  The GW-luminosity in the quadrupole
approximation is discussed in Sect.III while GW-amplitude with
gravitomagnetic corrections is discussed in Sect.IV giving also a
resume of numerical results.  Rate and event number estimations
are given in Sect.V. Conclusions are drawn in Sect.VI.

\section{Gravitomagnetic corrections}

In a previous paper \cite{Capoz}, we studied how the relativistic
theory of orbits for massive point-like  objects is affected by
gravitomagnetic corrections. In particular, we considered the
corrections on the orbits of higher-order terms in $v/c$ and this
is the main difference with respect to the standard
gravitomagnetic effect discussed so far where corrections are
taken into account only in the weak field limit and not on the
geodesic motion.   The problem of gravitomagnetic vector
potential, entering into the off-diagonal components $g_{0l}$ of
the metric $g_{\mu\nu}$, can be greatly simplified and the
corrections can be seen as  further powers in the expansion in
$c^{-1}$ (up to $c^{-3}$). Nevertheless, the effects on the orbit
behavior are interesting and involve not only the precession at
periastron but also nutation corrections. Here we briefly recall
such previous result sending to \cite{Capoz} for a detailed
analysis.

The metric, in weak field limit where   gravitomagnetic
corrections are present, is:

\begin{equation}
ds^{2}=\left(1+\frac{2\Phi}{c^{2}}\right)c^{2}dt^{2}-
\frac{8\delta_{lj}V^{l}}{c^{3}}cdtdx^{j}-\left(1-\frac{2\Phi}{c^{2}}\right)\delta_{lj}dx^{i}dx^{j}\;,
\label{eq:ds_DUE}\end{equation} where the gravitational Newtonian
potential $\Phi(x)$ is

\begin{equation}
\Phi(x)=-G\int\frac{\rho}{\left|\mathbf{x}-\mathbf{x}'\right|}d^{3}x'\;,\label{eq:fi_x}\end{equation}
and  the vector potential $V^{l}$ is

\begin{equation}
V^{l}=-G\int\frac{\rho
v^{l}}{\left|\mathbf{x}-\mathbf{x}'\right|}d^{3}x'\;,\label{eq:Vl}\end{equation}
given by the  matter current density $\rho v^{l}$ of the moving
bodies. This last potential gives rise to the gravitomagnetic
corrections.

It is clear that the approximation is up to $c^{-3}$ in the Taylor
expansion. From Eq.(\ref{eq:ds_DUE}),  it is straightforward to
construct a variational principle from which the geodesic equation
follows. Then we can derive the orbital equations. As standard, we
have

\begin{equation}
\ddot{x}^{\alpha}+\Gamma_{\mu\nu}^{\alpha}\dot{x}^{\mu}\dot{x}^{\nu}=0\;,\label{eq:geodedica_uno}\end{equation}

where  dot indicates the differentiation with respect to the
affine parameter. In order to put in evidence the gravitomagnetic
contributions, let us explicitly calculate the Christoffel symbols
at lower orders. By some straightforward calculations, one gets

\begin{equation}
\begin{array}{cl}
\Gamma^0_{00} &=0\\
\Gamma^0_{0j} &=\frac{1}{c^2}\frac{\partial\Phi}{\partial x^j} \\
\Gamma^0_{ij} &=-\frac{2}{c^3}\left(\frac{\partial V^i}{\partial x^j}+\frac{\partial V^j}{\partial x^i}\right) \\
\Gamma^k_{00} &= \frac{1}{c^2}\frac{\partial\Phi}{\partial x^k}\\
\Gamma^k_{0j} &=\frac{2}{c^3}\left(\frac{\partial V^k}{\partial x^j}-\frac{\partial V^j}{\partial x^k}\right) \\
\Gamma^k_{ij} &= -\frac{1}{c^2}\left(\frac{\partial \Phi}{\partial
x^j}\delta^k_i+\frac{\partial \Phi}{\partial
x^i}\delta^k_j-\frac{\partial \Phi}{\partial
x^k}\delta_{ij}\right)\end{array}
\end{equation}

In the approximation which we are going to consider, we are
retaining terms up to the orders $\Phi/c^2$ and $V^j/c^3$. It is
important to point out that we are discarding terms like
$(\Phi/c^4)\partial\Phi/\partial x^k$,
$(V^j/c^5)\partial\Phi/\partial x^k$, $(\Phi/c^5)\partial
V^k/\partial x^j$, $(V^k/c^6)\partial V^j/\partial x^i$ and of
higher orders. Our aim is to show that, in several  cases like in
tight binary stars, it is not correct to discard higher order
terms in $v/c$ since physically interesting effects could come
out. A vector  equation for the spatial components of geodesics
accounting for the gravitomagnetic effects is \cite{Capoz}:
\begin{equation}
\frac{d\mathbf{e}}{dl_{euclid}}=-\frac{2}{c^2}\left[\nabla\Phi-\mathbf{e}(\mathbf{e}\cdot\nabla\Phi)\right]+\frac{4}{c^3}
\left[\mathbf{e}\wedge(\nabla\wedge\mathbf{V})\right]\label{vector}\,.
\end{equation}
The gravitomagnetic term is the second in Eq.(\ref{vector}) and it
is usually discarded since considered not relevant. This is not
true if $v/c$ is quite large as in the cases of tight binary
systems or point masses approaching to black holes.   Orbits,
corrected by such effects can be explicitly achieved.

\subsection{Orbits with gravitomagnetic corrections}

Orbits with gravitomagnetic effects can be obtained starting from
the classical theory and then correcting it by successive
relativistic terms. In \cite{Capoz}, it is shown that, taking into
account the  gravitomagnetic terms, in the weak field
approximation and in the extreme mass-ratio limit $m \ll M$, one
obtains a motion with precession and nutation  by solving
numerically the Euler-Lagrange equations.  It is possible to
obtain the parametric orbital equations of a massive particle
starting from a variational principle where the canonical
Lagrangian is derived by the metric (\ref{eq:ds_DUE}). Being
$\frac{\partial\mathcal{L}}{\partial t}=0$, we have
$\frac{d}{dt}\left[\frac{\partial\mathcal{L}}{\partial\dot{t}}\right]=0$
and then $\frac{\partial\mathcal{L}}{\partial\dot{t}}=E$ where $E$
is a constant that can be interpreted as an energy per mass unit.
Owing to the dependence of the Lagrangian from $\theta$ and
$\phi$, we have, in general, $\frac{\partial\mathcal{L}}{\partial
\phi}\neq 0$ and, furthermore, considering the initial conditions
 $\theta=\frac{\pi}{2}$ and $\dot{\theta}=0$ we have
$\ddot{\theta}\neq 0$. This means, by straightforward
calculations,

\begin{eqnarray}
\ddot{\theta}=\frac{4 c^3 \text{E} G M r \dot{r}-16 \cos  G^2 M^2
\phi  \left(\dot{r}\right)^2-16 G^2 M^2 \sin  \phi
\left(\dot{r}\right)^2-16 \cos  G^2 M^2 r \phi  \dot{r}
\dot{\phi}+16 G^2 M^2 r \sin  \phi  \dot{r} \dot{\phi}}{r^2+c^6
r^4-4 c^2 G^2 M^2\,. } \label{thetaddot}
\end{eqnarray}

As it  is possible to see from Eq.(\ref{thetaddot}), planar motion
is possible setting the initial condition $\dot{r}=0$,  i.e. for
the particular case of circular orbits; otherwise  orbital motions
present precession and nutation corrections. Giving explicitly the
energy first integral

\begin{eqnarray}
 \dot{t}= \frac{1}{c^3 r-2 c G \mu }\left.\Big \{-4 G \mu  (\cos \theta +\sin \theta  (\cos\phi +\sin \phi )) \dot{r}  +r \left.\Big[c^3 E -4 G \mu  \left((-\sin \theta + \cos\theta (\cos \phi +\sin \phi )) \dot{\theta}\right.\right.\right.\\
\nonumber
 \left.\left. + \sin \theta (\cos \phi -\sin \phi ) \dot{\phi}\right.)\Big]
 \right.\Big\}\,,
\end{eqnarray}
and the Euler-Lagrange equations (where the  energy first integral
can be suitably substituted) we get the differential system
  \begin{eqnarray}
  \label{ddr}
 \ddot{r}= \frac{1}{c r \left(r c^2+2 G \mu\right)}
 \Big[c \left(r c^2+G \mu\right) \left(\dot{\theta}^2+\sin ^2\theta \dot{\phi}^2\right) r^2\\
 \nonumber
 -4 G \mu \dot{t} \left((\cos\theta(\cos\phi+\sin\phi)-\sin \theta)\dot{\theta}+
 \sin\theta (\cos\phi-\sin \phi)\dot{\phi} \right) r+c G \mu \dot{r}^2-c G \mu
 \dot{t}^2\Big]\,,
  \end{eqnarray}

 \begin{eqnarray}
 \label{ddphi}
\ddot{\phi}=-\frac{2 \left(c \cot \theta \left(r c^2+2 G \mu\right) \dot{\theta}\dot{ \phi} r^2+\dot{r}
   \left(2 G \mu \csc \theta (\sin\phi-\cos\phi) \dot{t}+c r \left(r c^2+G
   \mu\right) \dot{\phi}\right)\right)}{r^2 \left(r c^3+2 G \mu
   c\right)}\,,
   \end{eqnarray}

  \begin{eqnarray}
  \label{ddtheta}
 \ddot{\theta}=\frac{c \cos\theta r^2 \left(r c^2+2 G \mu\right) \sin\theta \dot{\phi}^2+\dot{r}
   \left(4 G \mu (\cos\theta (\cos \phi+\sin\phi)-\sin\theta)\dot{t}-2 c r \left(r c^2+G \mu\right)
   \dot{\theta}\right)}{r^2 \left(r c^3+2 G \mu c\right)}\,.
    \end{eqnarray}

\begin{figure}[!ht]
\begin{tabular}{|c|c|}
\hline
\tabularnewline
\includegraphics[scale=0.4]{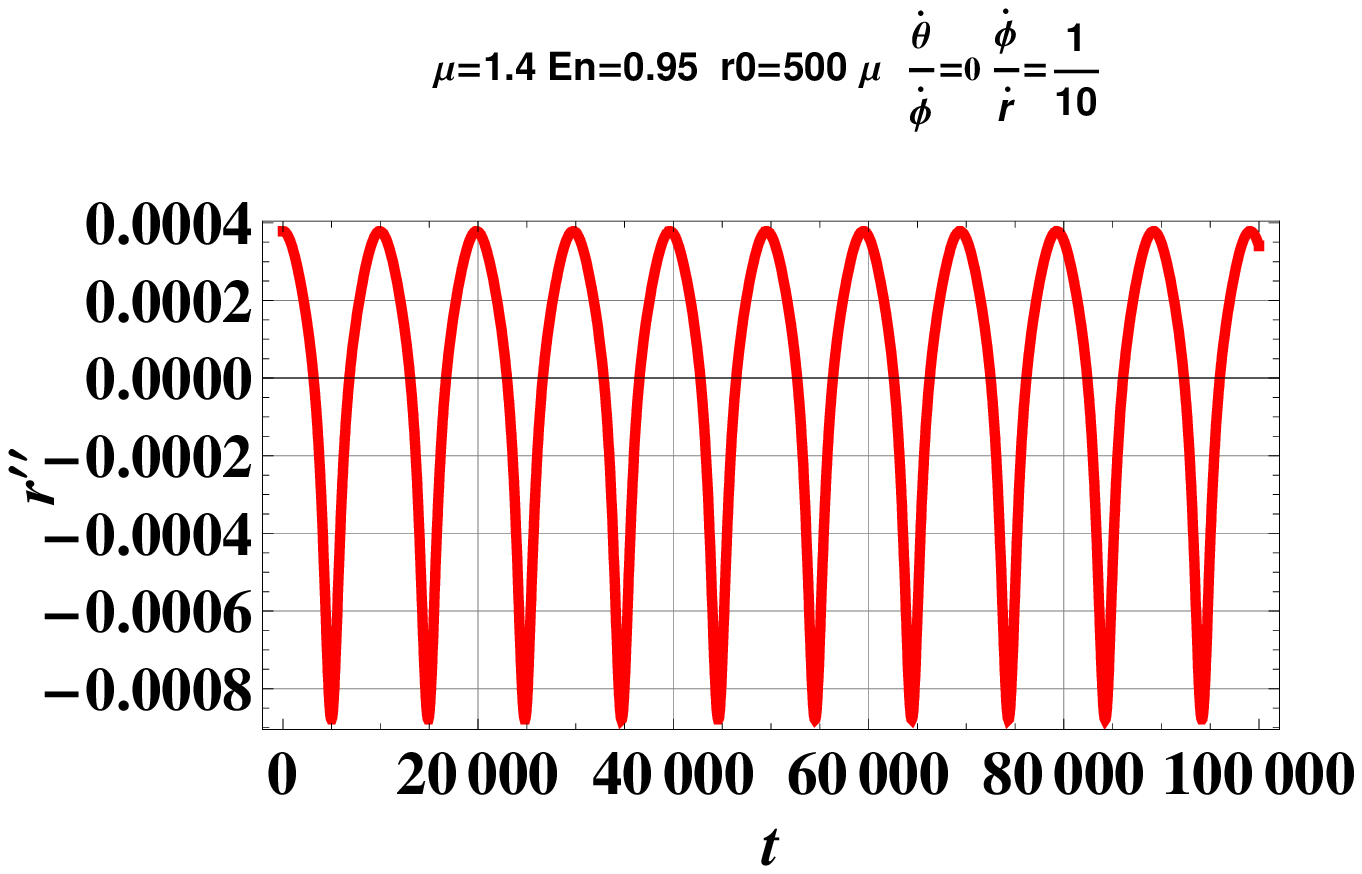}
\includegraphics[scale=0.4]{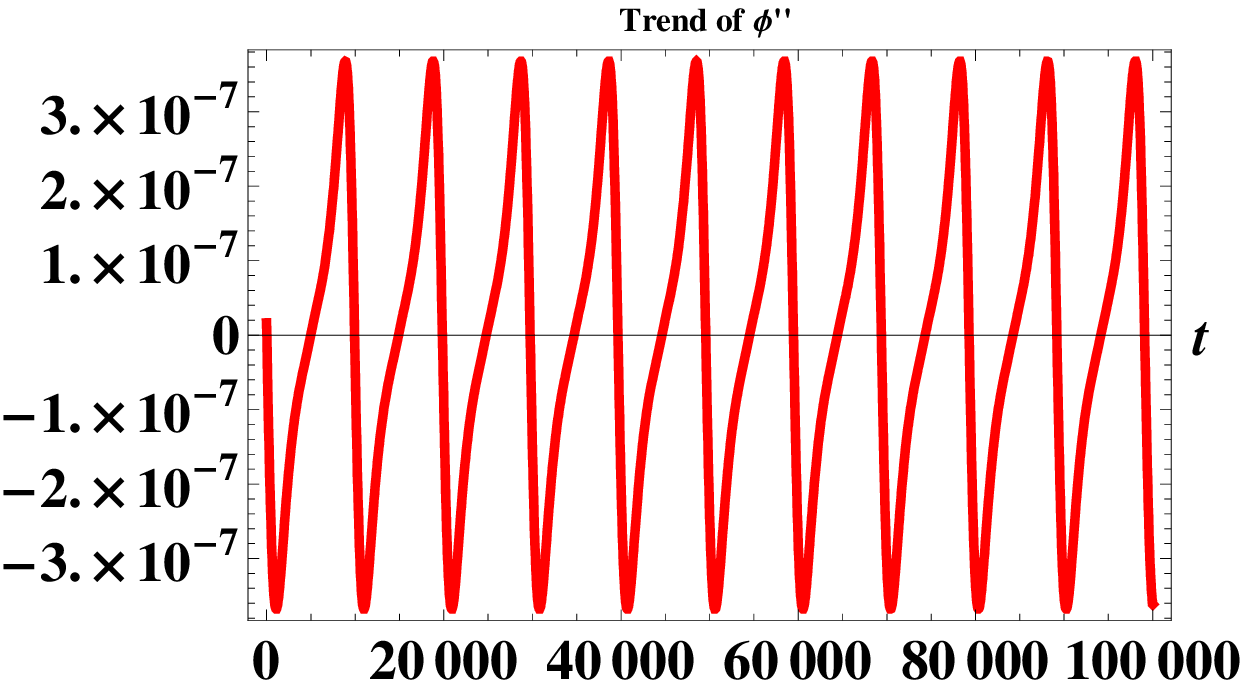} \tabularnewline
\hline
\includegraphics[scale=0.4]{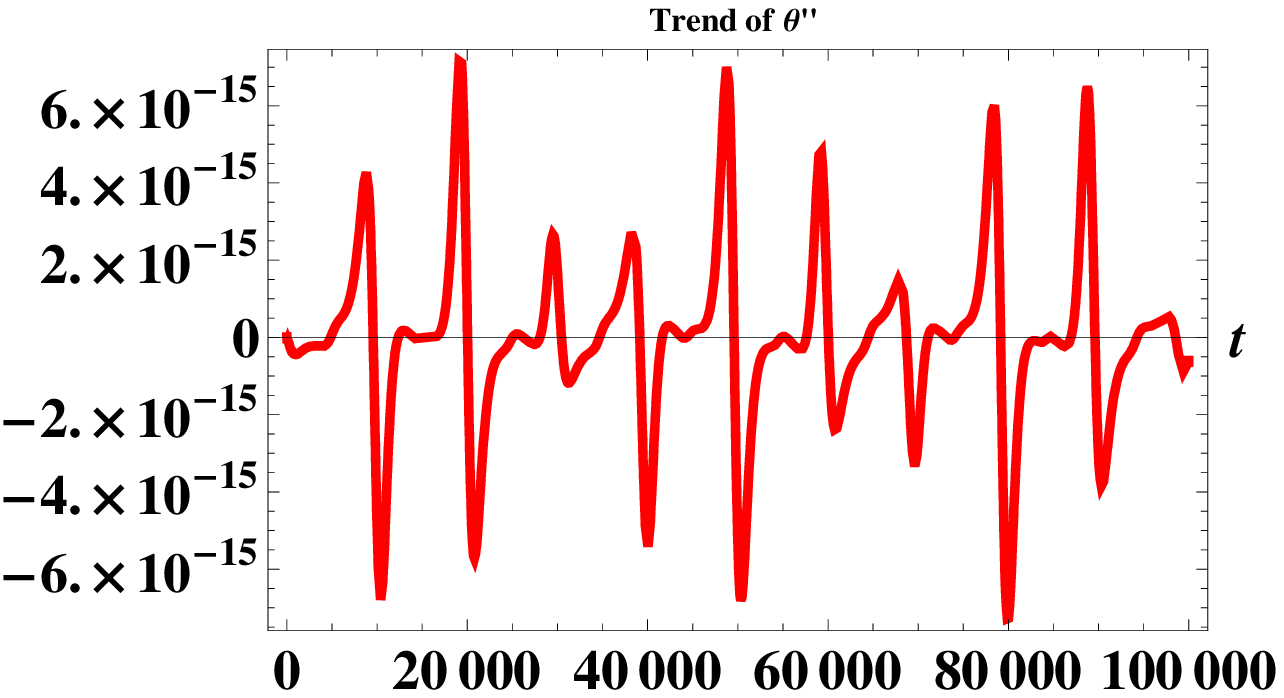}
\includegraphics[scale=0.4]{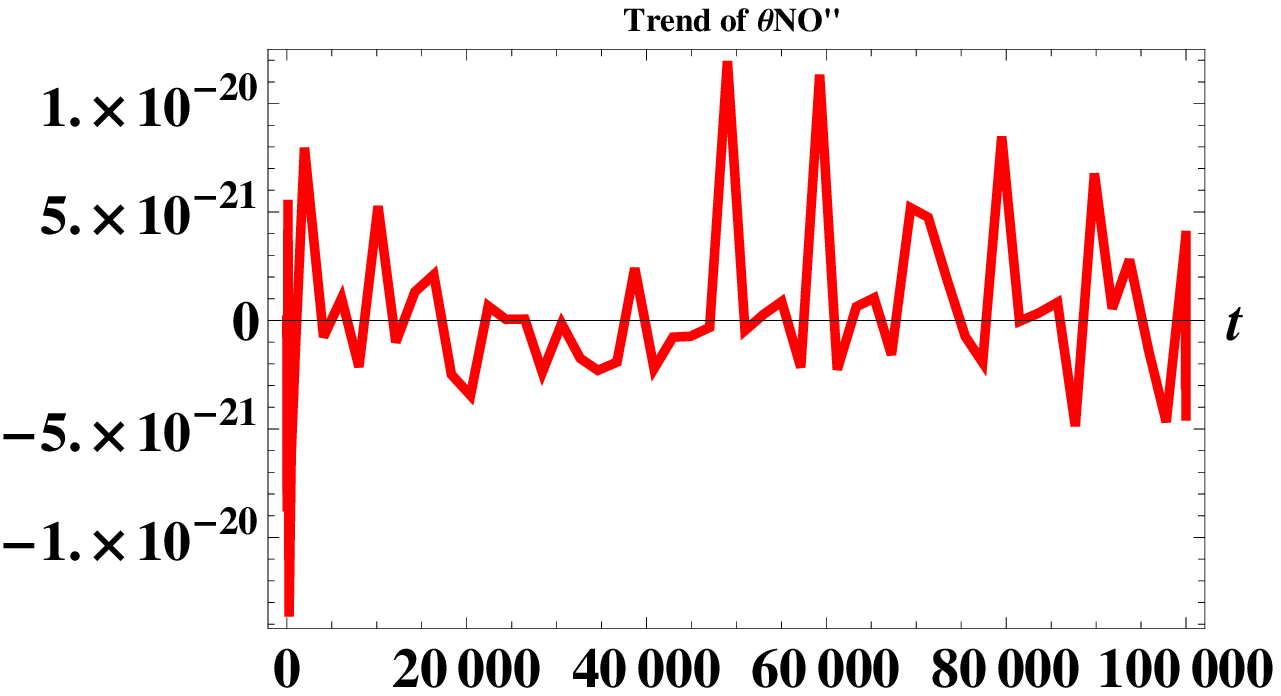}
\tabularnewline
\hline
\end{tabular}
\caption {Plots of $\ddot{r}=\ddot{r}(t)$(left upper
panel),$\ddot{\phi}=\ddot{\phi}(t)$ (right upper panel),
$\ddot{\theta}=\ddot{\theta}(t)$(left bottom
panel),$\ddot{\theta}_{NO}=\ddot{\theta}_{NO}(t)$  (right bottom
panel). As it is possible to see we have stiff equations owing to
the turning points of the orbits. The example we are showing has
been obtained solving the system for the following parameters and
initial conditions: $\mu=1.4 M_{\odot}$, $r_{0}=500\mu$, $E=0.95$,
$\phi_{0}=0$, $\theta_{0}=\frac{\pi}{2}$,
$\dot{\phi_{0}}=-\frac{1}{10}\dot{r}_{0}$ and
$\dot{r}_{0}=-\frac{1}{100}$.}\label{Fig:01}
\end{figure}

Such a system is  highly non linear. For its solution,  we have to
adopt numerical methods. The main characteristics of solutions can
be evaluated seeing at Fig.\ref{Fig:01} where the stiffness of the
system is evident.

From Eqs.(\ref{ddr})-(\ref{ddtheta}), it is clear that additional
terms appear with respect to  the classical Newtonian motion. Such
corrections  are not independent from $\phi$ and $\theta$.
Obviously, these terms become important as soon as velocities
approach relativistic regimes and the ratio ${\displaystyle
\frac{v}{c}}$ is quite large. In some physical situations, e.g. in
extreme dense globular clusters or around the Galactic Center,
such a ratio can be in the range $10^{-2}\div 10^{-3}$ being not
negligible at all (see \cite{freitag} and references therein).

In Fig.\ref{Fig:01}, we have shown the trend of $\ddot{r}$,
$\ddot{\phi}$, $\ddot{\theta}$, as function of $t$, and
$\ddot{\theta}_{NO}$ which is the trend without gravitomagnetic
correction. It can be seen that this last plot gives deviation
from zero that are essentially null, confirming the planarity of
orbital motions in absence of gravitomagnetic corrections (the
differences have at least four orders of magnitude between
$\ddot{\theta}_{NO}(t)$ and $\ddot{\theta}(t)$).

\begin{figure}[!ht]
\begin{tabular}{|c|c|}
\hline \tabularnewline
\includegraphics[scale=0.4]{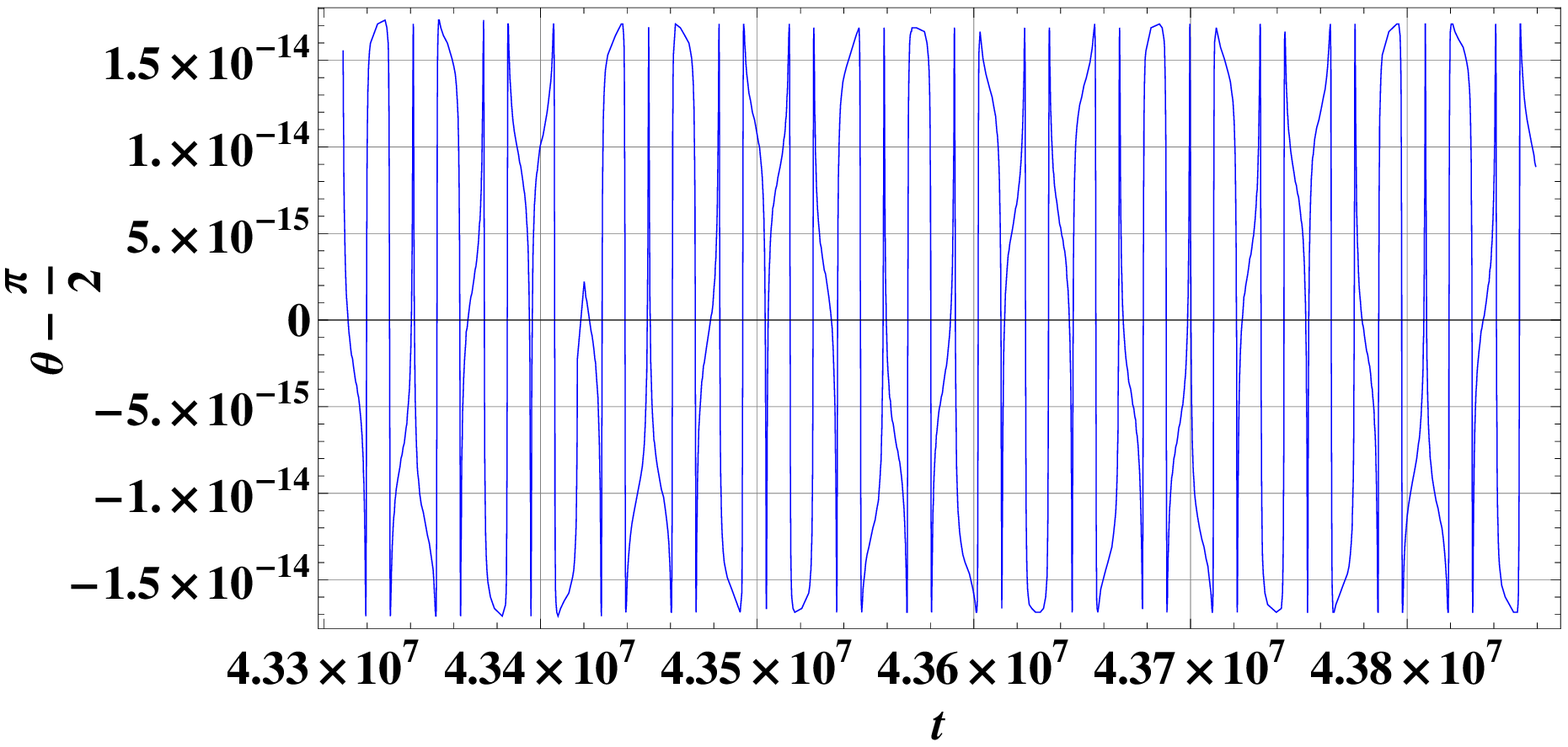}
\includegraphics[scale=0.4]{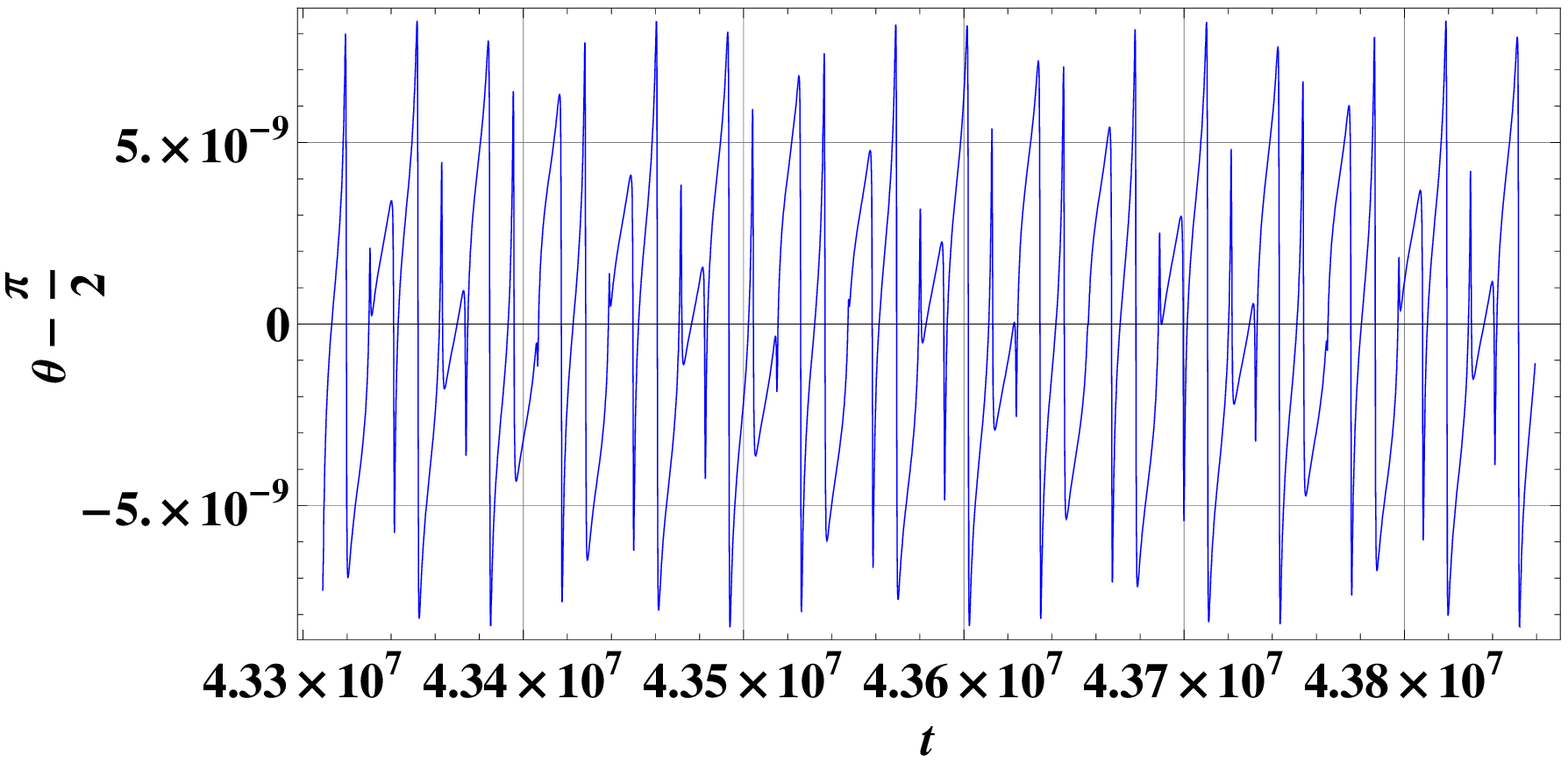}
\tabularnewline \hline
\includegraphics[scale=0.4]{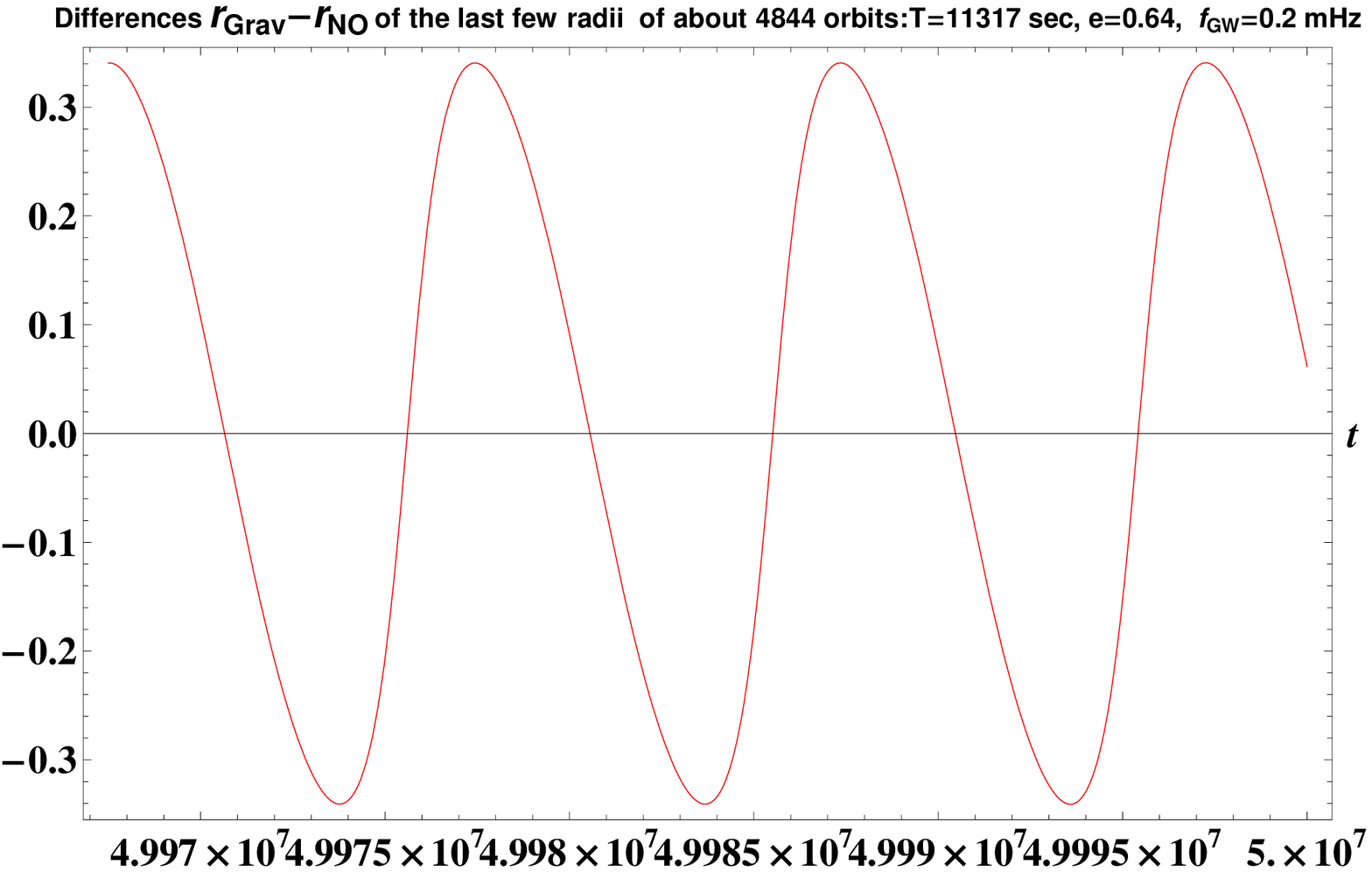}
\includegraphics[scale=0.4]{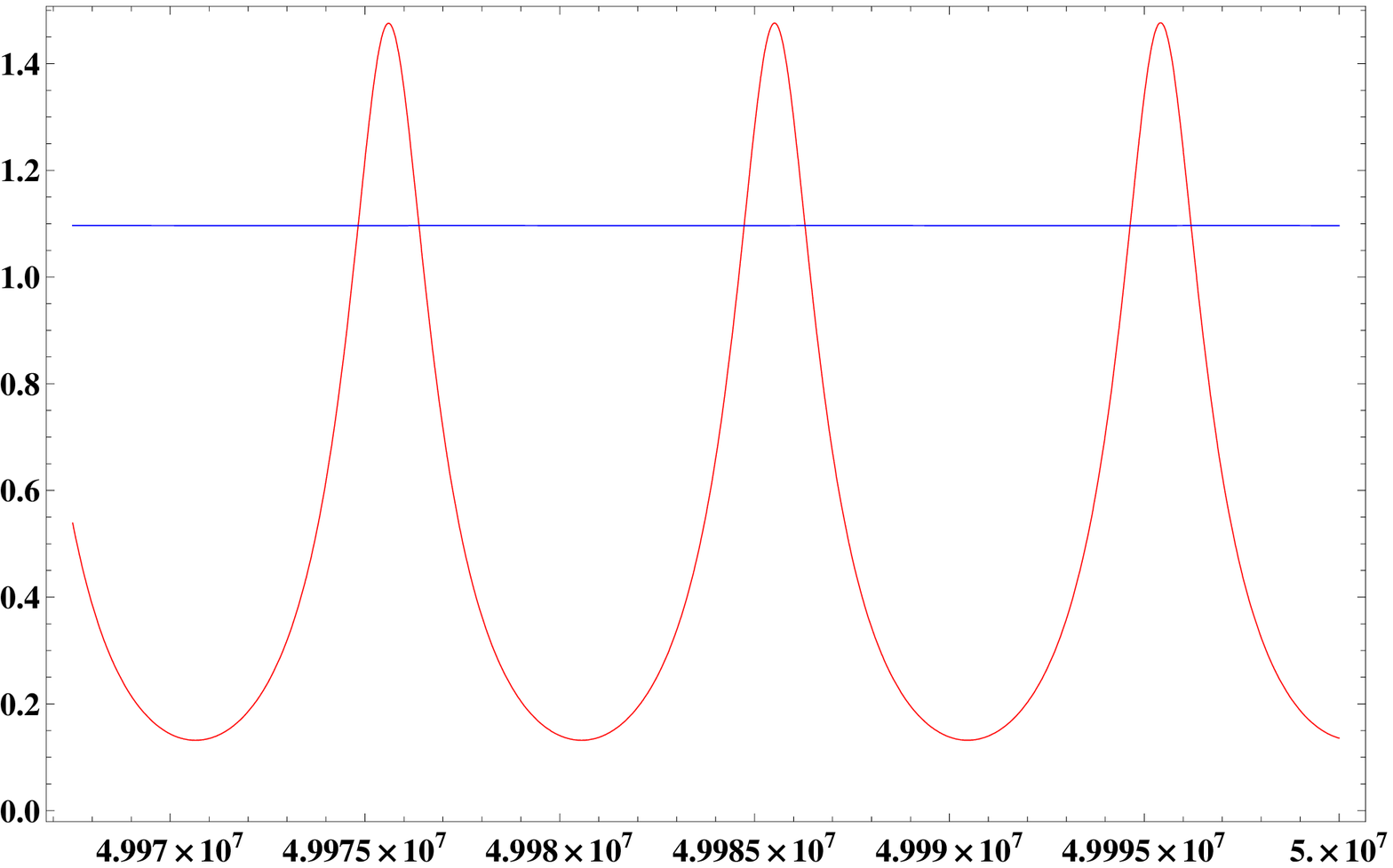}
\tabularnewline \hline
\end{tabular}
\caption {Plots  of $\theta_{NO}(t)-\frac{\pi}{2}$ (left upper
panel) and of $\theta_{Grav}(t)-\frac{\pi}{2}$ (right  upper
panel). In the bottom panels,  $r_{Grav}-r_{NO}$ (left) and
$t_{Grav}-t_{NO}$ (right) are plotted (red lines). The ratio
between coordinated time $\frac{t_{Grav}}{\tau}$ versus proper
time $\tau$ is also plotted (blue line). The examples we are
showing have been obtained solving the system for the following
parameters and initial conditions: $\mu=1.4 M_{\odot}$,
$r_{0}=500\mu$, $E=0.95$, $\phi_{0}=0$,
$\theta_{0}=\frac{\pi}{2}$, $\dot{{\theta}_{0}}=0$,
$\dot{\phi_{0}}=-\frac{1}{10}\dot{{r}_{0}}$ and
$\dot{r}_{0}=-\frac{1}{100}$.}\label{Fig:02}
\end{figure}

\begin{figure}[!ht]
\begin{tabular}{|c|c|}
\hline
\tabularnewline
\includegraphics[scale=0.5]{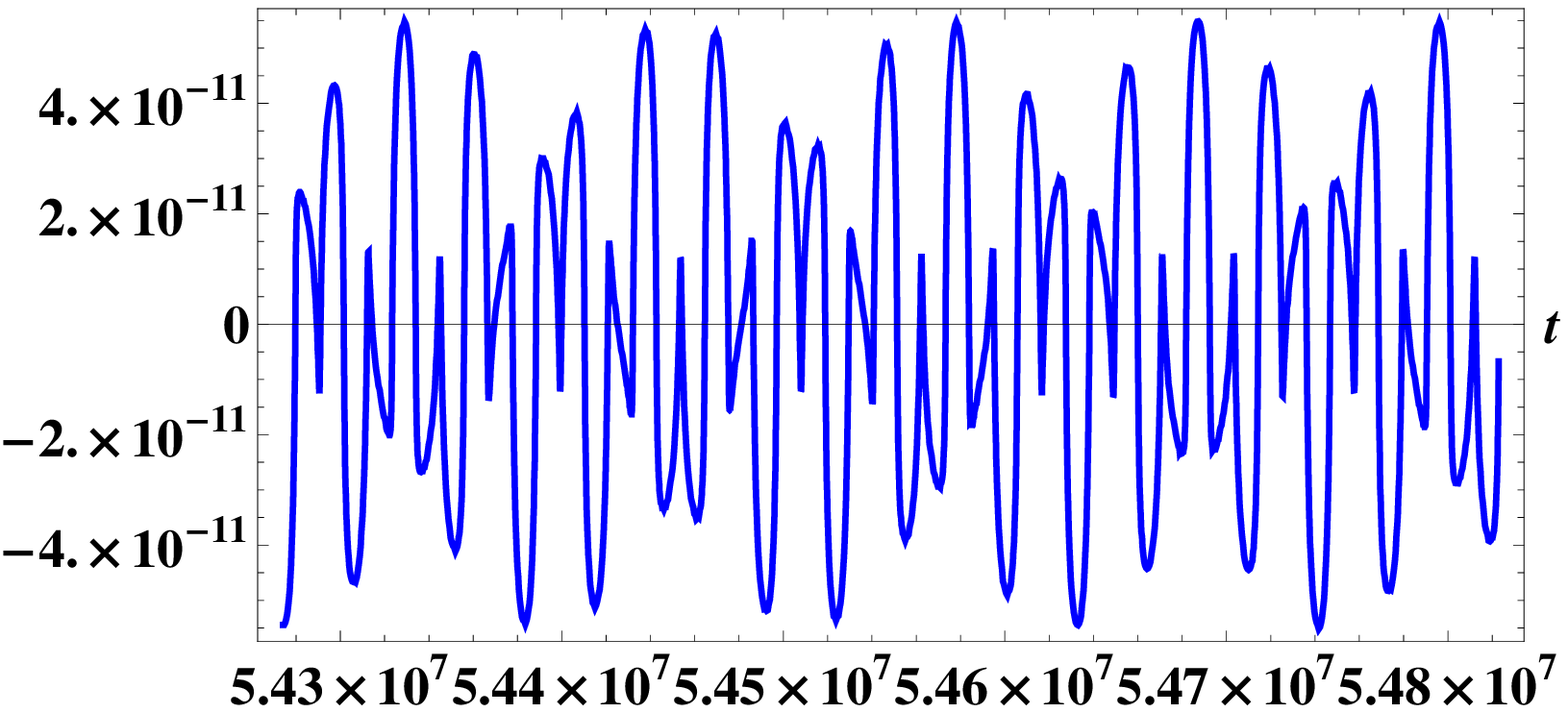}
\includegraphics[scale=0.5]{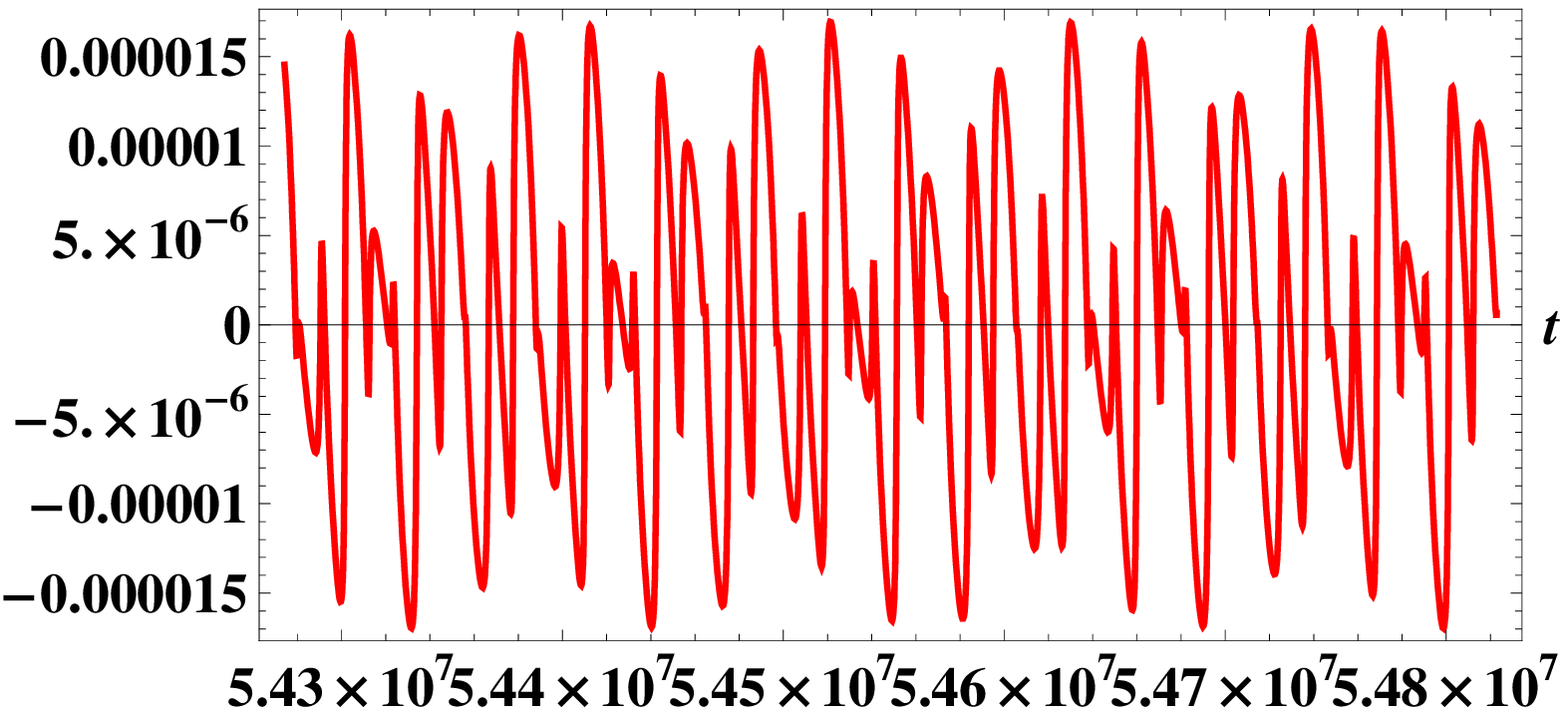}
\tabularnewline
\hline
\includegraphics[scale=0.5]{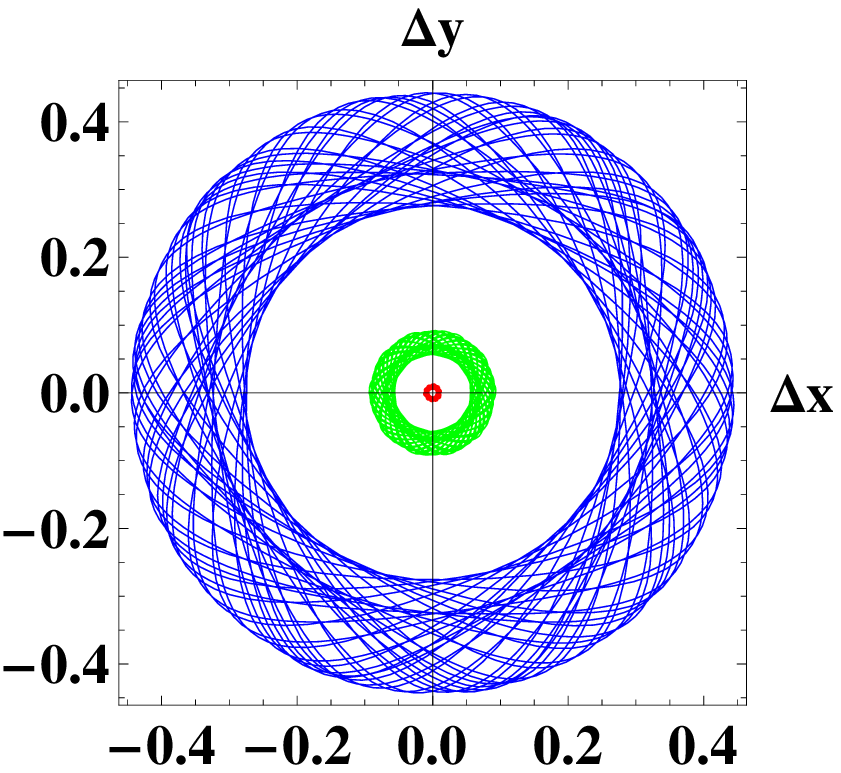}
\includegraphics[scale=0.5]{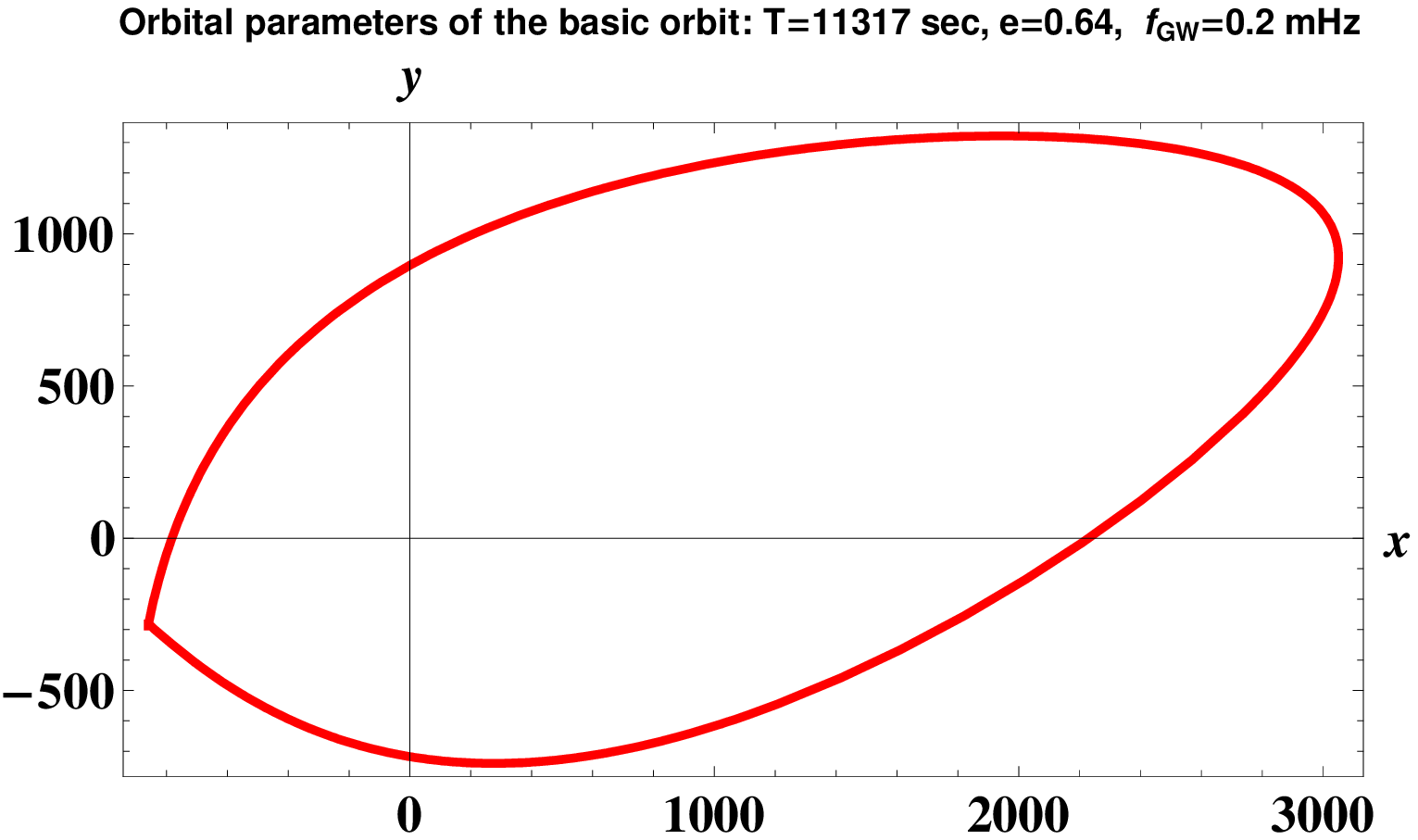}
\tabularnewline
\hline
\end{tabular}
\caption {Plots of $z_{NO}(t)$ (left upper panel) and
$z_{Grav}(t)$ (right upper panel). It is interesting to see the
differences of about five orders of magnitude between the two
plots. At the beginning, the effect is very small but, orbit by
orbit, it grows and, for a suitable interval of coordinated time,
the effect cannot be neglected (see the left bottom panel in which
the differences in $x$ and $y$, starting from the initial orbits
up to the last ones, by steps of about 1500 orbits, are reported).
The internal red circle represents the beginning, the middle one
is the intermediate situation (green) and the blue one is the
final result of the correlation between $\Delta x$ versus $\Delta
y$, being $\Delta x=x_{Grav}-x_{NO}$ and $\Delta
y=y_{Grav}-y_{NO}$. On the bottom right, it is shown the basic
orbit.}\label{Fig:03}
\end{figure}

\begin{figure}[!ht]
\begin{tabular}{|c|c|c|}
\hline
\tabularnewline
\includegraphics {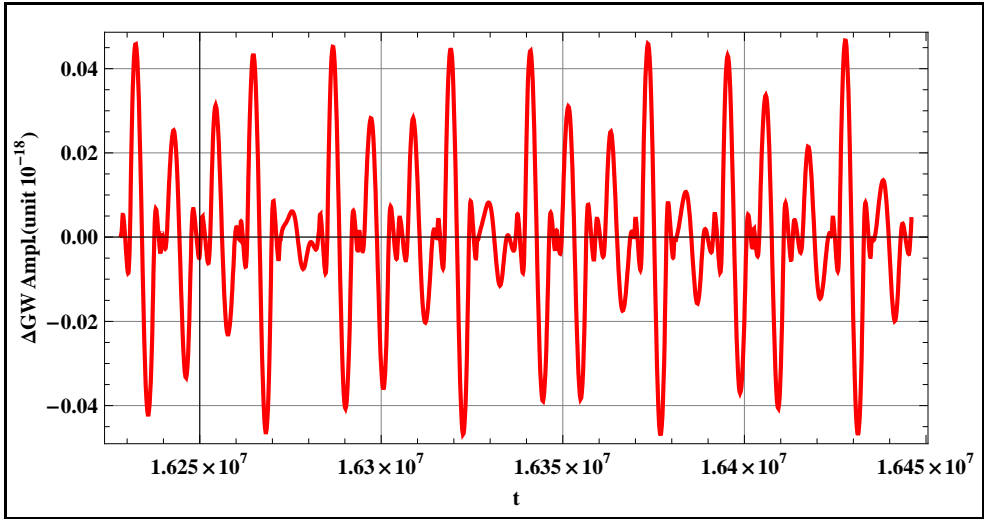}
 \tabularnewline
\hline
\end{tabular}
\caption {Plot of the differences of total gravitational waveform
$h$, with and without the gravitomagnetic orbital correction  for
a neutron star of $1.4 M_{\odot}$ orbiting around a MBH . The
waveform has been  computed at the Earth-distance from SgrA$^*$
(the central Galactic Black Hole). The example we are showing has
been obtained solving the systems for the following parameters and
initial conditions: $\mu\approx1.4 M_{\odot}$, $r_{0}$, $E=0.95$,
$\phi_{0}=0$, $\theta_{0}=\frac{\pi}{2}$, $\dot{\theta}_{0}=0$,
$\dot{\phi_{0}}=-\frac{1}{10}\dot{r}_{0}$ and
$\dot{r}_{0}=-\frac{1}{100}.$ It is worth noticing that frequency
modulation gives cumulative effects after suitable long times.
}\label{Fig:04}
\end{figure}

To have a further insight of the   gravitomagnetic correction
relevance on the relativistic orbital motion, we have derived a
numerical solution with the following parameters and initial
conditions: ${\displaystyle \mu=\frac{m M}{m+M}\cong 1.4
M_{\odot}}$, $r_{0}=500\mu$, $E=0.95$, $\phi_{0}=0$,
$\theta_{0}=\frac{\pi}{2}$,
$\dot{\phi_{0}}=-\frac{1}{10}\dot{r}_{0}$ and
$\dot{r}_{0}=-\frac{1}{100}c$. In Fig.(\ref{Fig:02}), we have
plotted   $\theta_{NO}(t)-\frac{\pi}{2}$  without gravitomagnetic
corrections  and $\theta_{Grav}(t)-\frac{\pi}{2}$, with
gravitomagnetic corrections. In the bottom panel, there is,
starting from the left to the right, the trend of the difference
between the orbital radii $r_{Grav}$ and  $r_{NO}$ with and
without  gravitomagnetic corrections respectively; we plotted also
the differences $r_{Grav}-r_{NO}$ and  $t_{Grav}-t_{NO}$(red
lines) and the ratio between coordinated time ${\displaystyle
\frac{t_{Grav}}{\tau}}$ versus proper time $\tau$ (blue line). It
is interesting to see the discrepancy from $\frac{\pi}{2}$ of
$\theta$  with and without the gravitomagnetic effect. It is
evident that we have planar orbital motion in the Newtonian case,
whilst, in presence of  gravitomagnetic corrections, there is a
tendency to precession and nutation of the orbital plane which
give rise, orbit by orbit, to cumulative effects  (a difference of
five orders of magnitude between $z_{NO}(t)$ and $z_{Grav}(t)$ can
be evaluated). At the beginning, the effect is very small but,
orbit by orbit, it grows and, for a suitable interval of  time,
the effect cannot be neglected [see Fig.(\ref{Fig:03}), left
bottom panel in which the differences in $x$ and $y$ are shown
starting from the initial orbits up to the last ones by step of
about 1500 orbits). On the bottom right, it is shown the basic
orbit. For about 4850 orbits and a time interval of about $1.7$
years,  we found that the differences in coordinated time,
computed with and without gravitomagnetic effects, are increasing
as well as the differences in $x$, $y$ and $z$ coordinates. See
also Fig. (\ref{Fig:04}) in which we show the differences between
the gravitational wave strain amplitude computed with and without
the gravitomagnetic orbital corrections (see the discussion in
forthcoming Sect. IV).

\begin{figure}[!ht]
\begin{tabular}{|c|c|c|}
\hline
\tabularnewline
\includegraphics[scale=0.25]{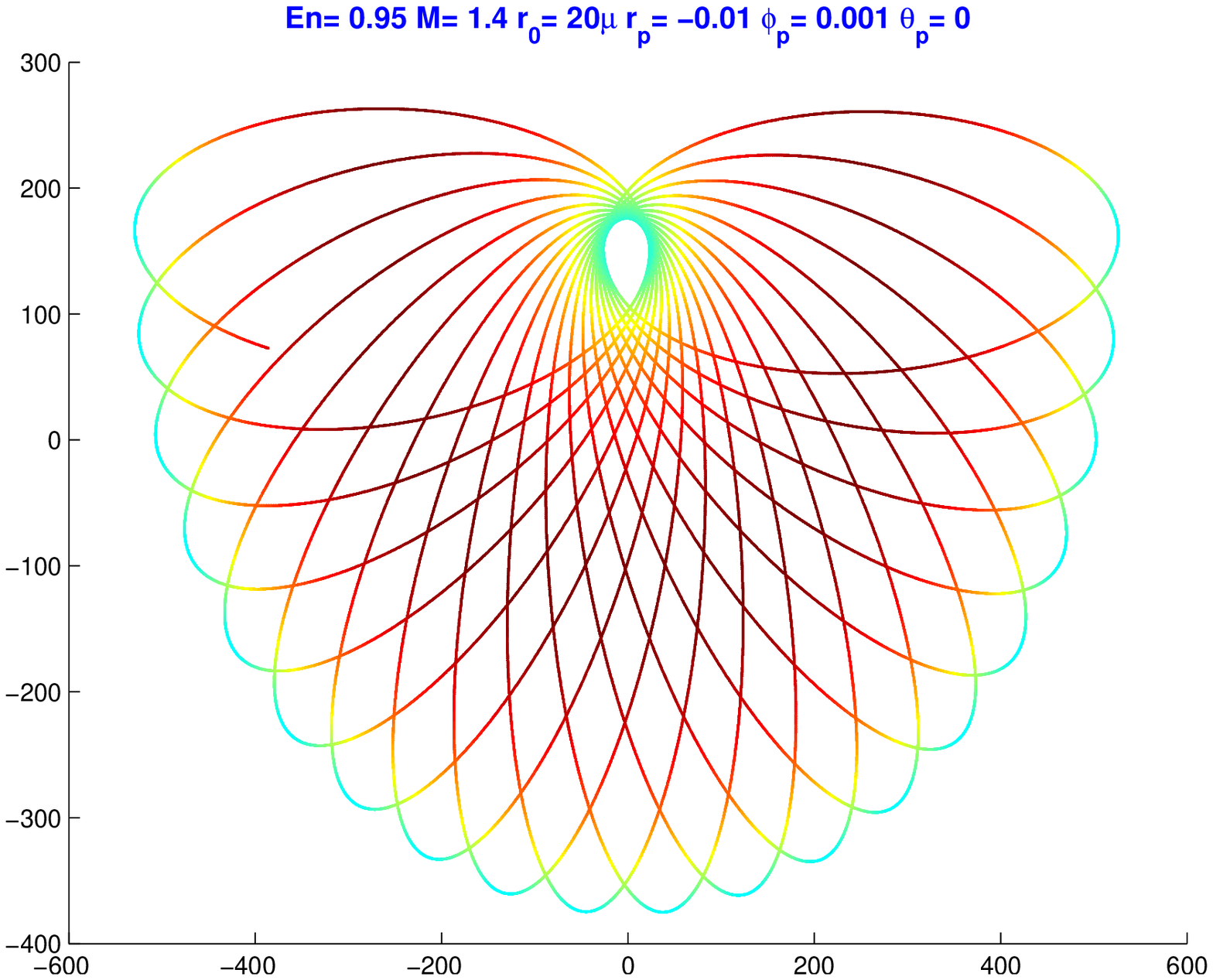}
\includegraphics[scale=0.25]{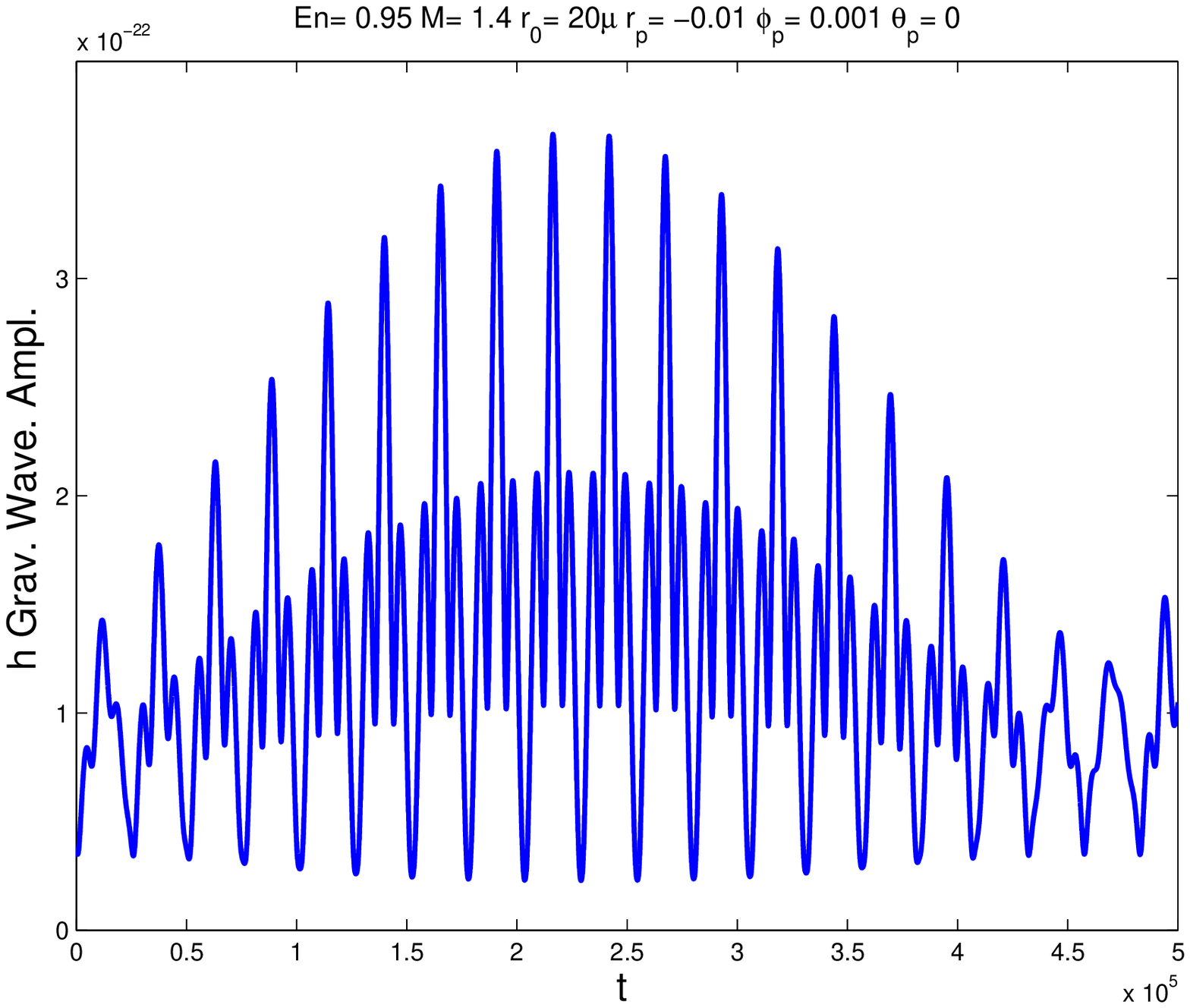}
\includegraphics[scale=0.25]{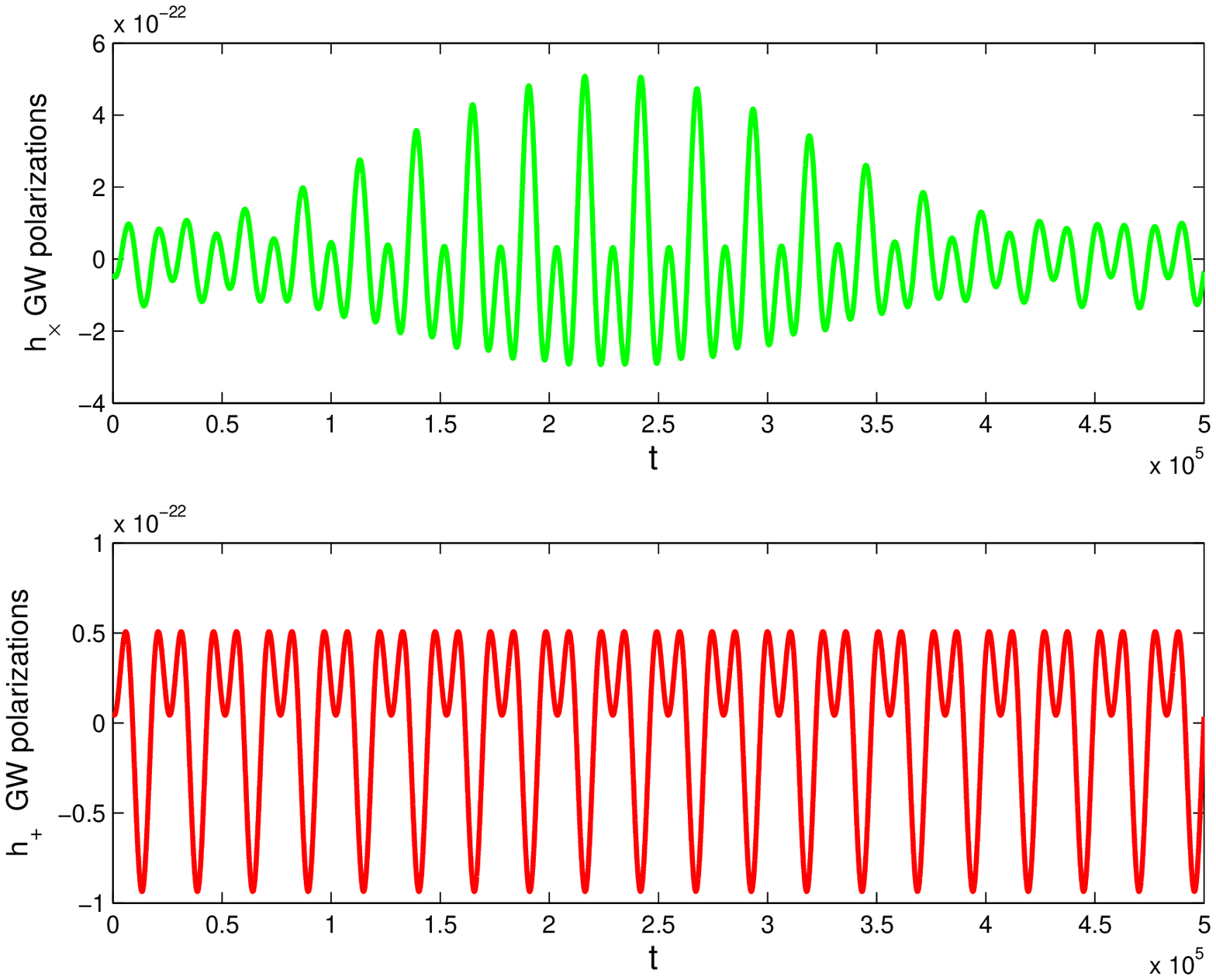}
 \tabularnewline
\hline
 \tabularnewline
\includegraphics[scale=0.25]{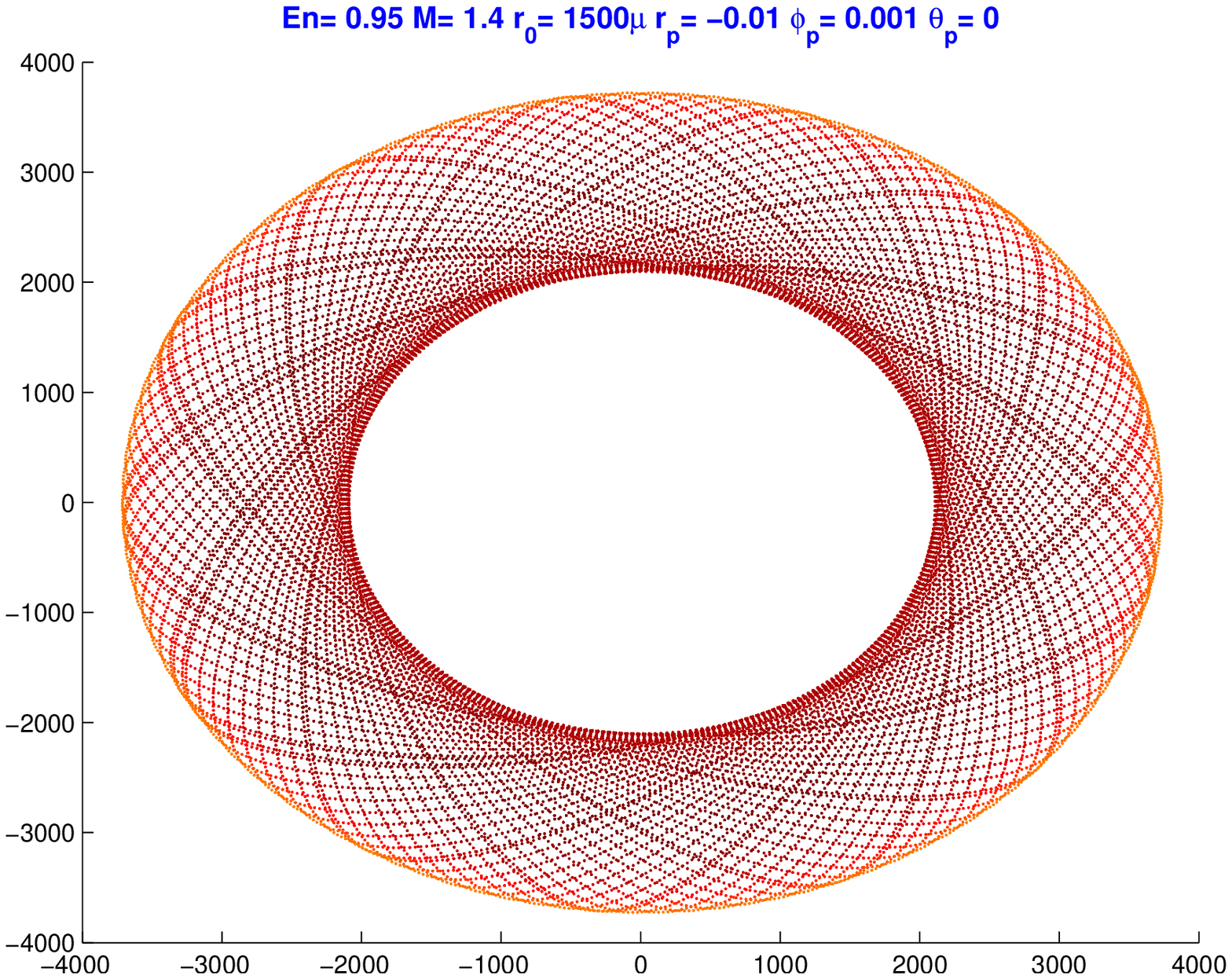}
\includegraphics[scale=0.25]{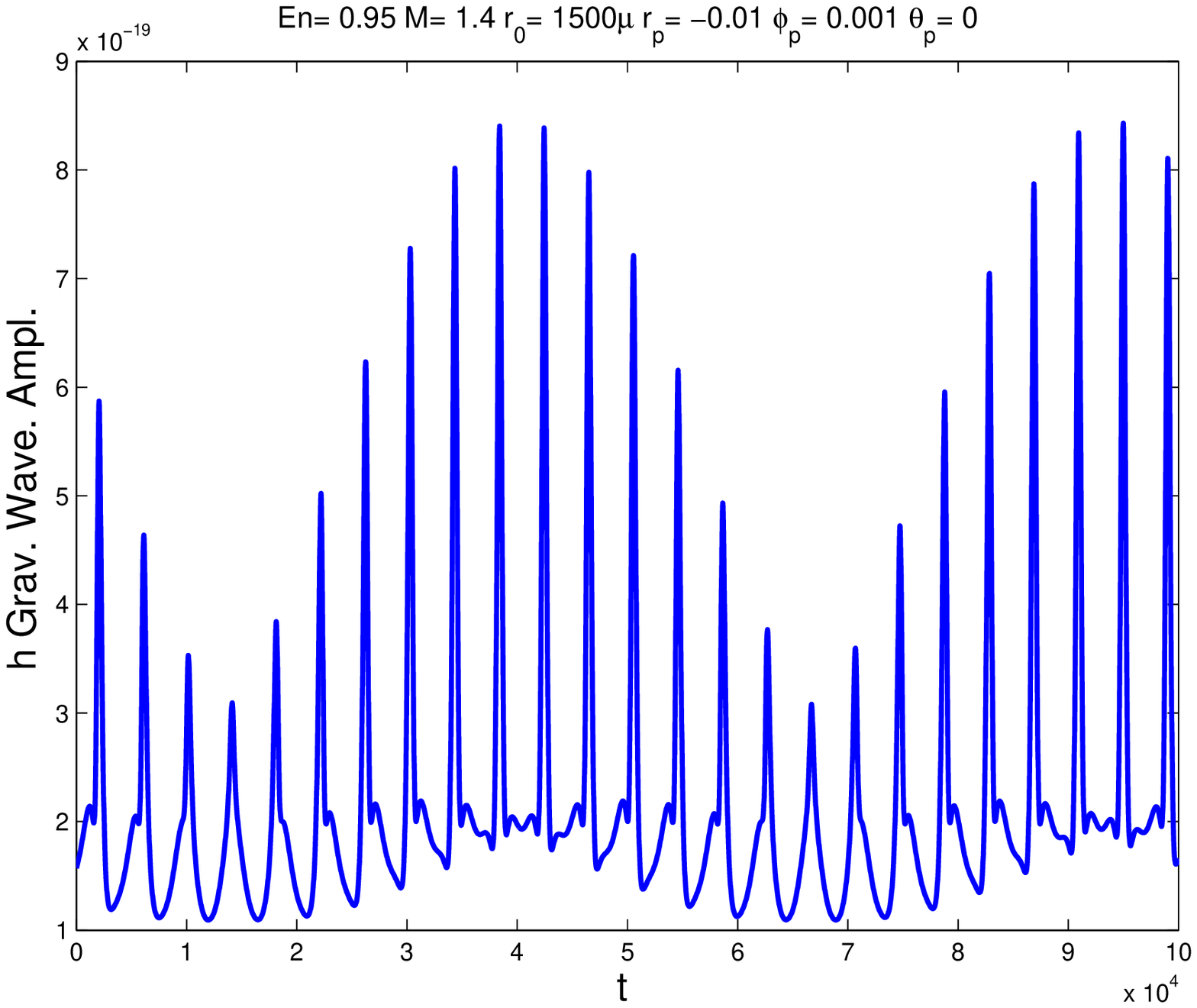}
\includegraphics[scale=0.25]{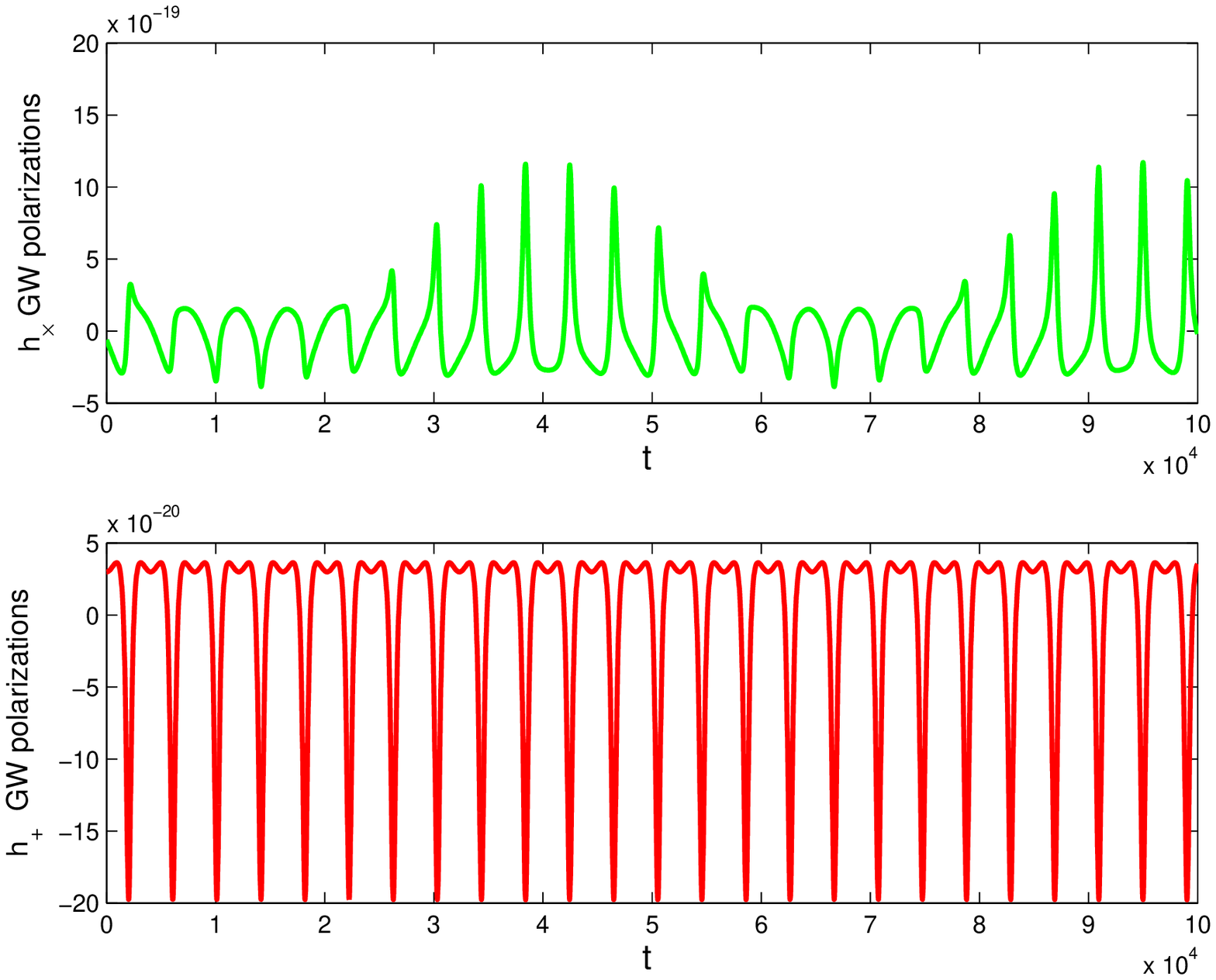}
\tabularnewline
\hline
\tabularnewline
\includegraphics[scale=0.25]{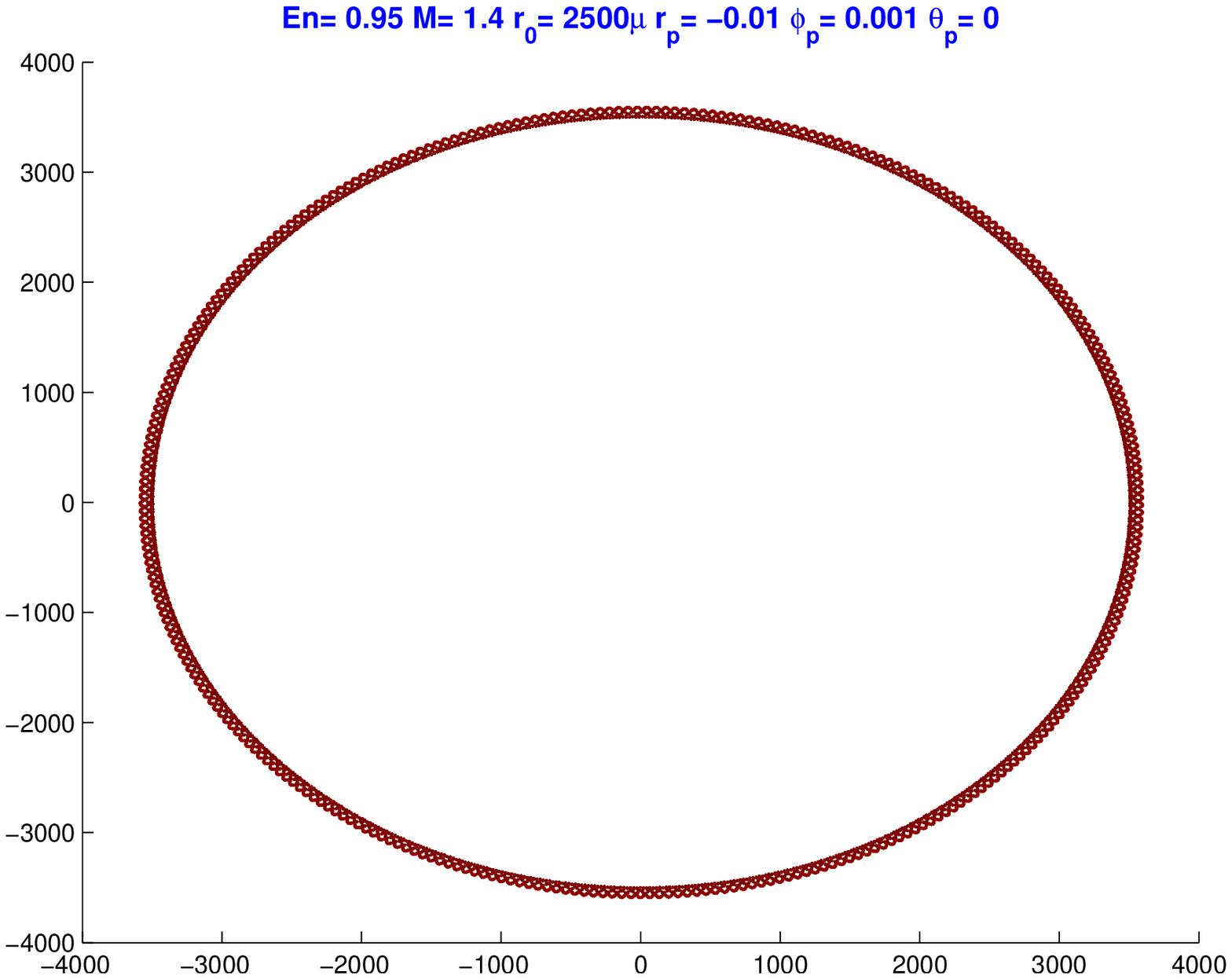}
\includegraphics[scale=0.25]{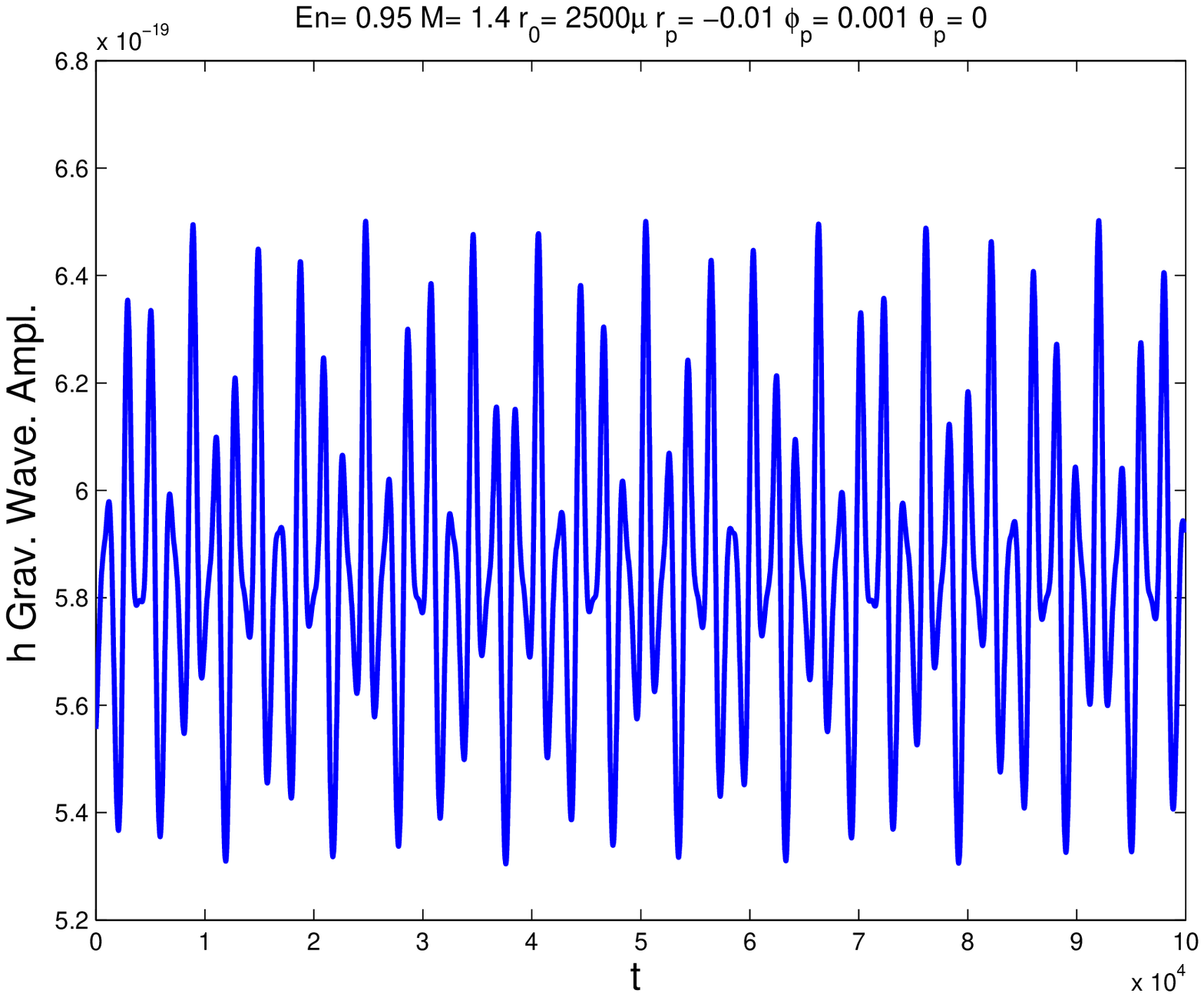}
\includegraphics[scale=0.25]{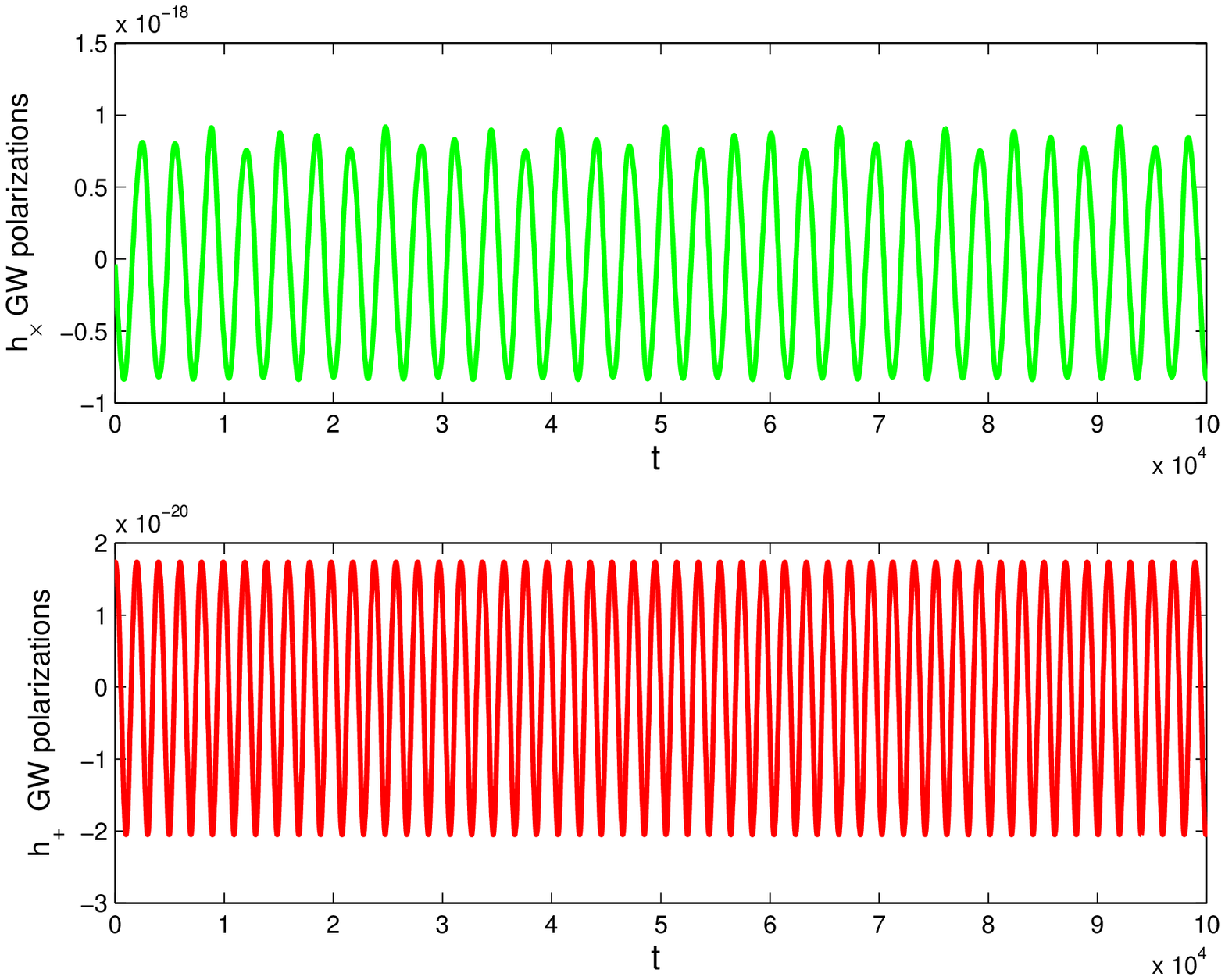}
 \tabularnewline
\hline
\end{tabular}
\caption {Plots along the panel  lines from left to right of field
velocities along the axes of maximum covariances, total
gravitational emission waveform $h$ and gravitational waveform
polarizations $h_{+}$ and $h_{\times}$ for a neutron star of $1.4
M_{\odot}$. The waveform has been computed for the Earth-distance
from Sagittarius A (the central Galactic Black Hole SgrA$^*$). The
plots we are showing have been obtained solving the system for the
following parameters and initial conditions: $\mu\approx1.4
M_{\odot}$, $E=0.95$, $\phi_{0}=0$, $\theta_{0}=\frac{\pi}{2}$,
$\dot{\theta_{0}}=0$, $\dot{\phi_{0}}=-\frac{1}{10}\dot{r}_{0}$
and $\dot{r}_{0}=-\frac{1}{100}$. From top to bottom of the
panels,  the  orbital radius is $r_0=20\mu,\,1500\mu,\,2500\mu$.
See also Table I.}\label{Fig:05}
\end{figure}

\begin{figure}[!ht]
\begin{tabular}{|c|c|c|}
\hline
\tabularnewline
\includegraphics[scale=0.25]{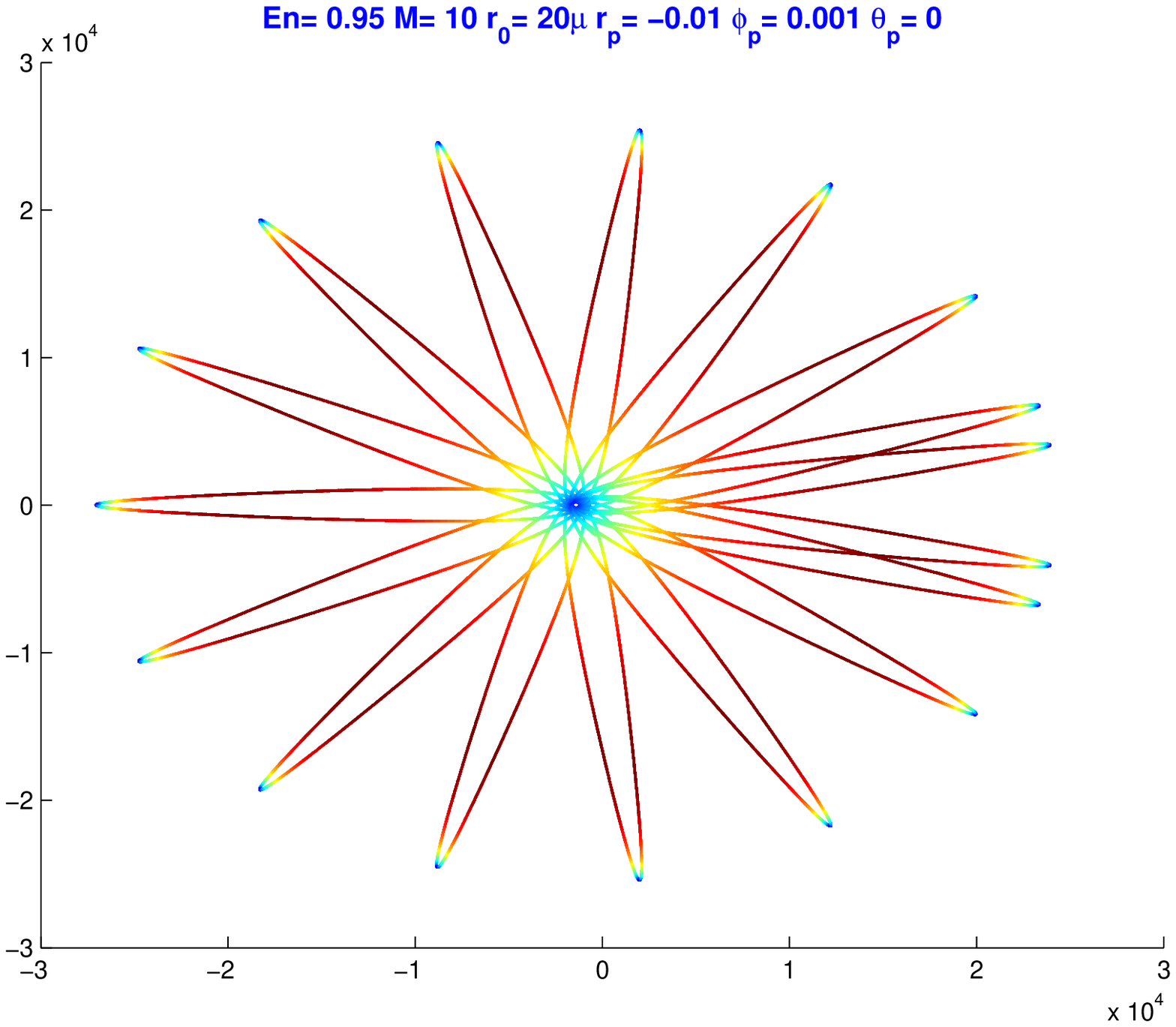}
\includegraphics[scale=0.25]{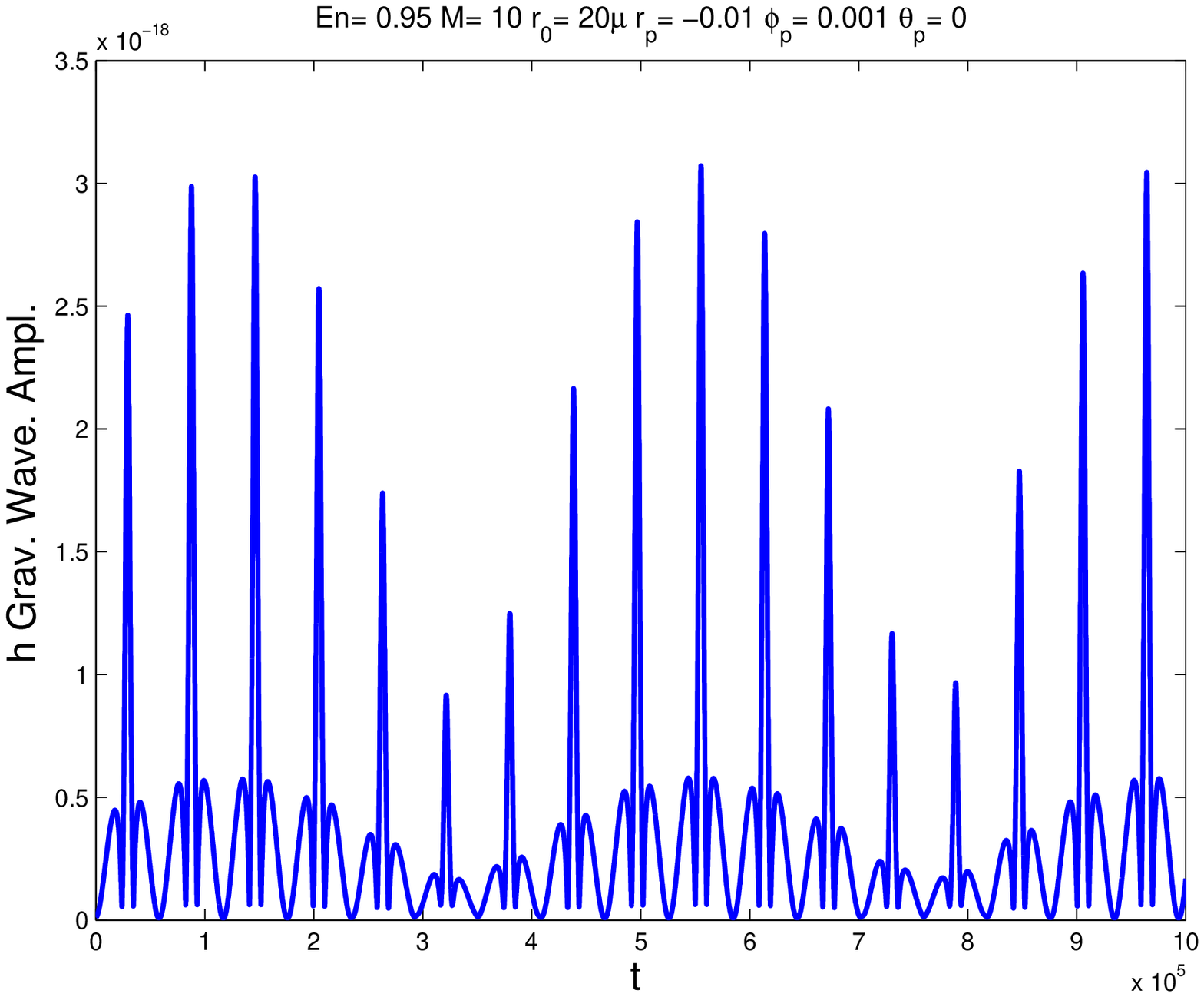}
\includegraphics[scale=0.25]{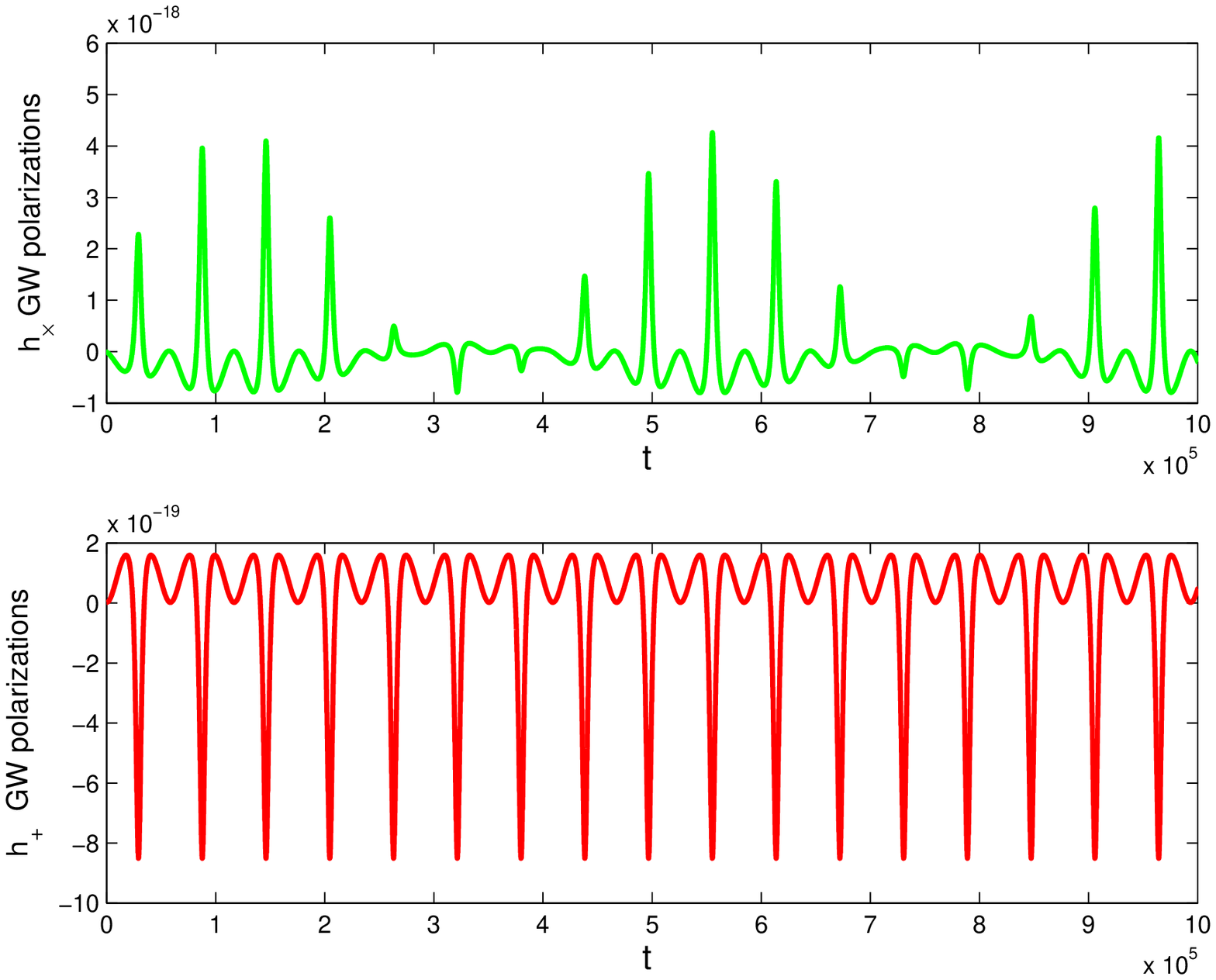}
 \tabularnewline
\hline
\includegraphics[scale=0.25]{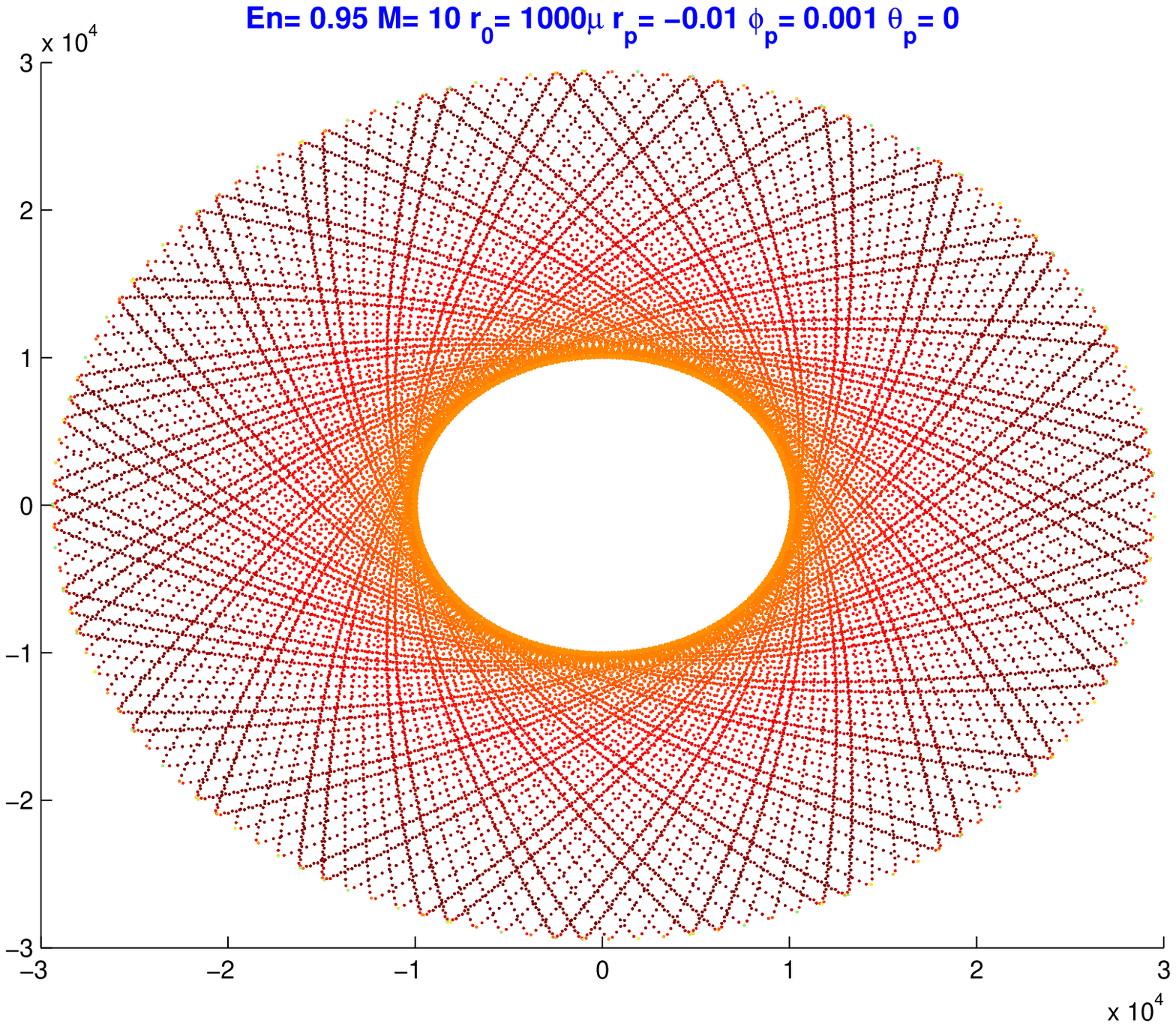}
\includegraphics[scale=0.25]{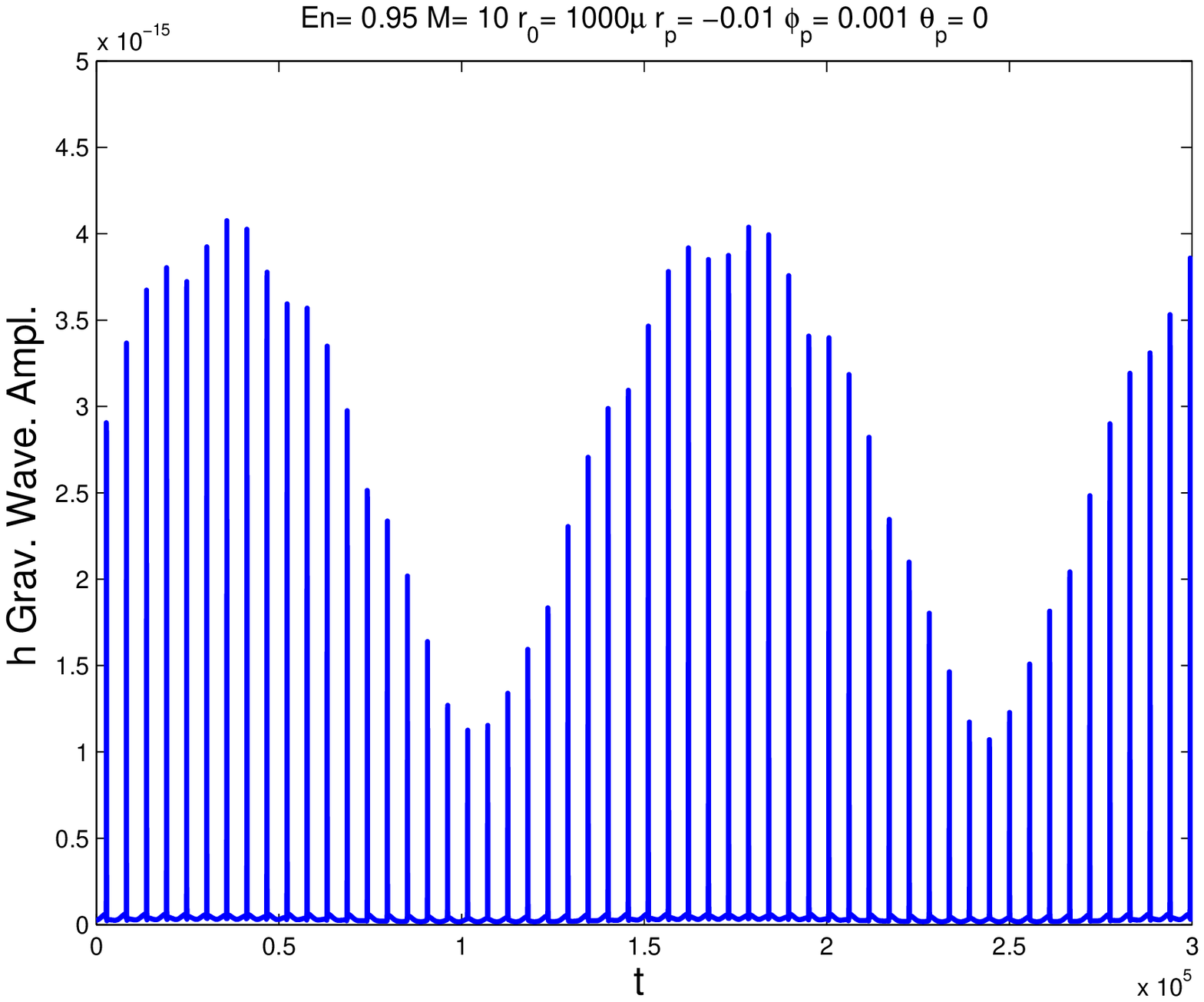}
\includegraphics[scale=0.25]{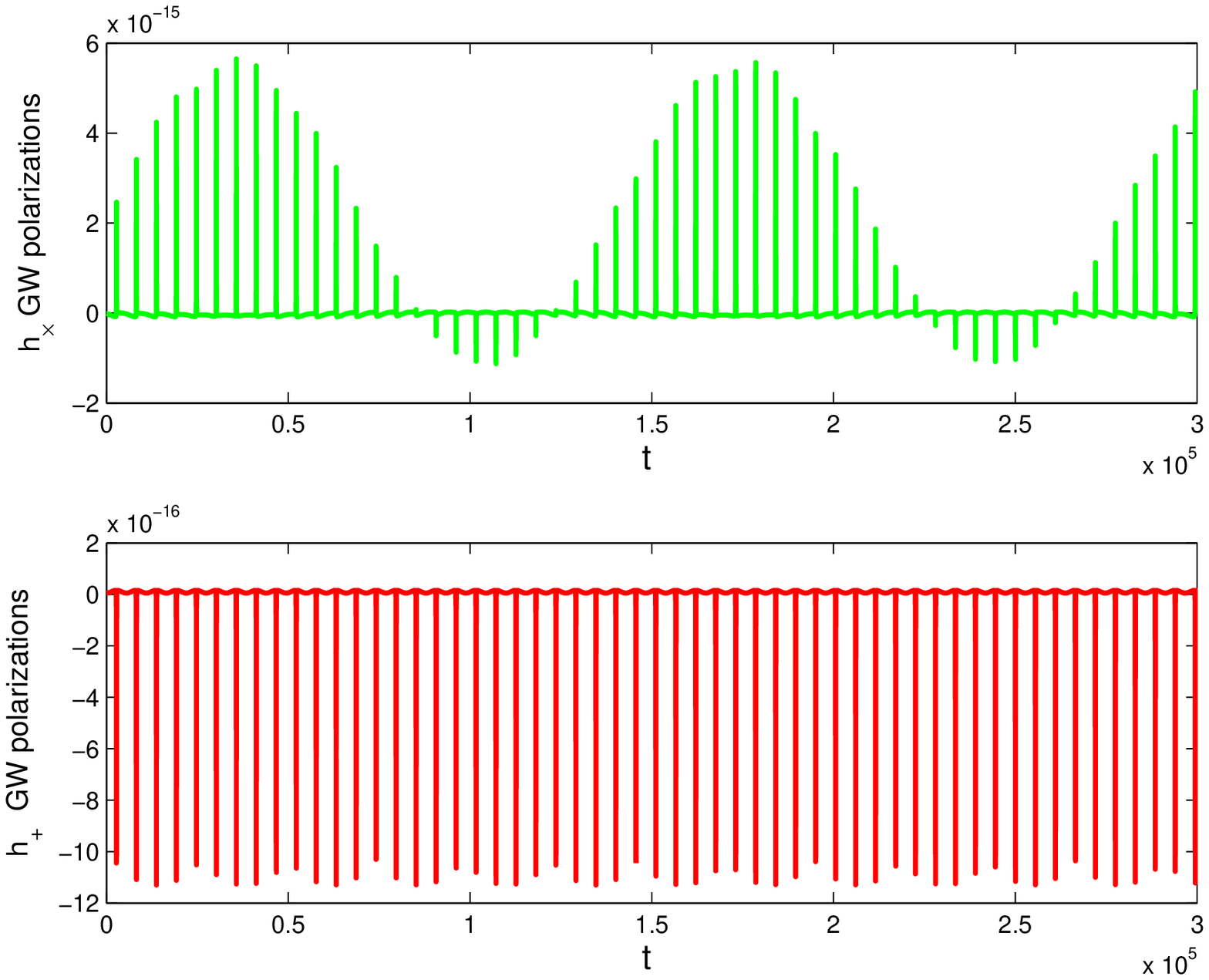}
\tabularnewline
\hline
 \tabularnewline
\includegraphics[scale=0.25]{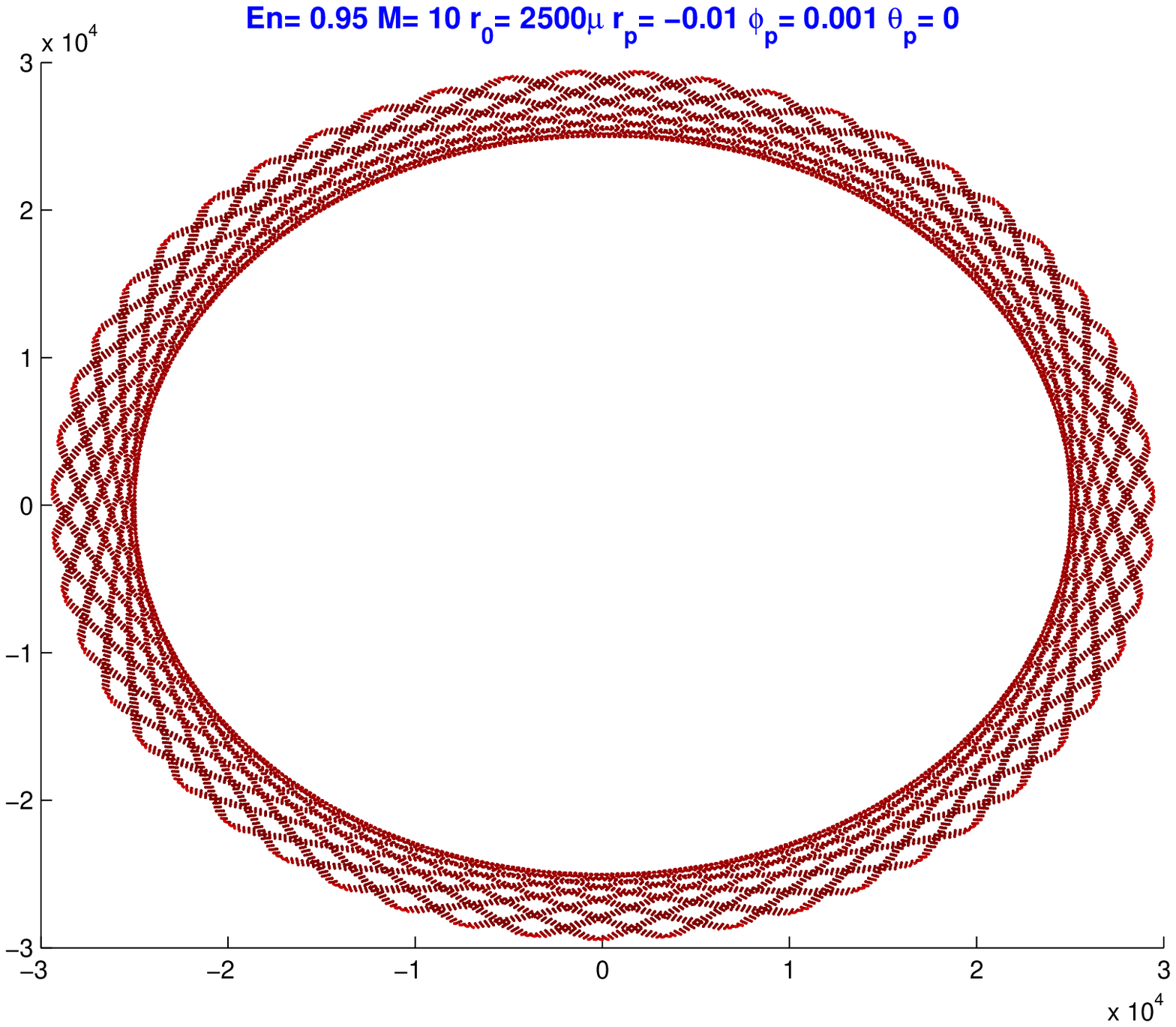}
\includegraphics[scale=0.25]{En_95_M_1.4_r0_2500_theta_p_0quadQ.eps}
\includegraphics[scale=0.25]{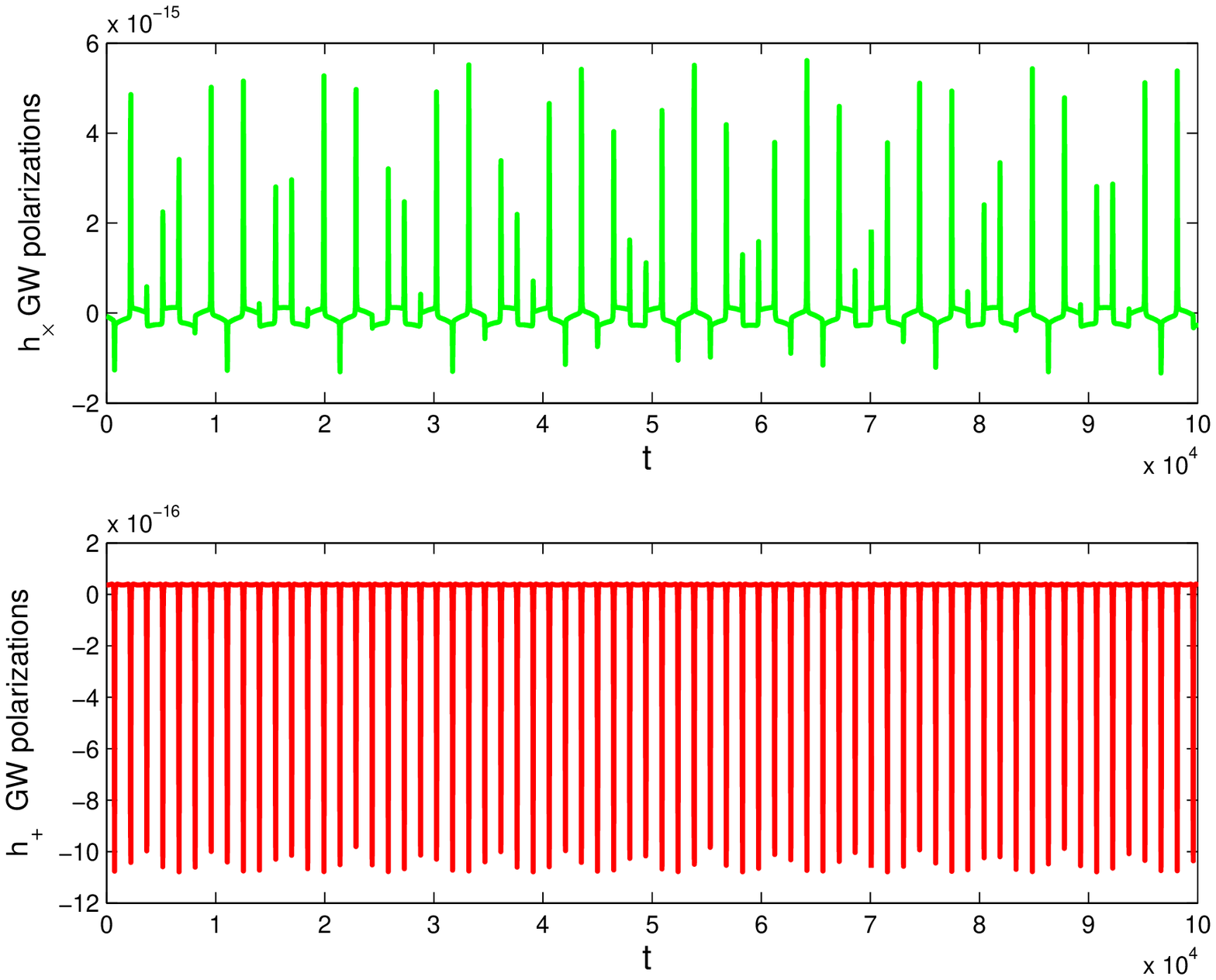}
\tabularnewline
\hline
\end{tabular}
\caption {Plots along the  panel lines from left to right of field
velocities along the axes of maximum covariances, total
gravitational emission waveform $h$ and gravitational waveform
polarizations $h_{+}$ and $h_{\times}$ for a Black Hole (BH) of
$10 M_{\odot}$. The waveform has been computed for the
Earth-distance to SgrA$^*$. The plots we are showing have been
obtained solving the system for the following parameters and
initial conditions: $\mu\approx10 M_{\odot}$,
$E=0.95$,$\phi_{0}=0$,
$\theta_{0}=\frac{\pi}{2}$,$\dot{\theta_{0}}=0$,$\dot{\phi_{0}}=-\frac{1}{10}\dot{r}_{0}$
and $\dot{r}_{0}=-\frac{1}{100}$. From top to bottom of the
panels, the  orbital radius is $r_0=20\mu,\,1000\mu,\,2500\mu$.
See also Table I}\label{Fig:07}
\end{figure}

\section{Gravitational wave luminosity in the quadrupole
approximation}

After the discussion of gravitomagnetic corrections on the orbital
motion, let us take into account the problem of how GW production
and waveforms are affected by such effects. To this purpose, let
us start with a short review of the quadrupole approximation for
the gravitational radiation. This is, in our opinion,  the best
way to see how gravitomagnetic effects correct GW luminosity and
waveforms.

It is well known that the Einstein field equations give a
description of how the curvature of space-time is related to the
energy-momentum distribution. In the weak field approximation,
moving massive objects produce gravitational waves which propagate
in the vacuum with the speed of light. In this approximation, we
have

\begin{equation}
g_{\mu\nu}=\delta_{\mu\nu}+\kappa h_{\mu\nu},\qquad\left(\left|h_{\mu\nu}\right|<<1\right),
\end{equation}
being $\kappa$ the gravitational coupling. The field equations are

\begin{equation}
 \square\bar{h}_{\mu\nu}=-\frac{1}{2}\kappa
T_{\mu\nu}\label{eq:h}\end{equation} where

\begin{equation}
\bar{h}_{\mu\nu}=h_{\mu\nu}-\frac{1}{2}\delta_{\mu\nu}h_{\lambda\lambda}\,,\end{equation}

and $T_{\mu\nu}$ is the total stress-momentum-energy tensor of the
source, including the gravitational stresses.

A plane GW, solution of (\ref{eq:h}), can be written as

\begin{equation}
\label{wave} \bar{h}_{\mu\nu}=h_{\mu\nu}=he_{\mu\nu}\cos(\omega
t-\boldsymbol{\mathbf{k}\cdot\mathbf{x}})
\end{equation}
where $h$ is the amplitude, $\omega$ the frequency, $k$ the wave
number and $e_{\mu\nu}$ is a unit polarization tensor, obeying the
conditions

\begin{equation}
e_{\mu\nu}=e_{\nu\mu},\qquad e_{\mu\mu}=0,\qquad
e_{\mu\nu}e_{\mu\nu}=1.\end{equation}
Assuming a  gauge in which
$e_{\mu\nu}$ is space-like and transverse, a wave travelling in
the \emph{z} direction has two possible independent polarizations:

\begin{equation}
e_{1}=\frac{1}{\sqrt{2}}(\hat{x}\hat{x}-\hat{y}\hat{y})\qquad
e_{2}=\frac{1}{\sqrt{2}}(\hat{x}\hat{y}-\hat{y}\hat{x}).\end{equation}
One can  search for wave solutions of (\ref{eq:h}), generated by a
system of masses undergoing arbitrary motions, and then obtain the
radiated power. The result, assuming the source dimensions very
small with respect to the wavelengths (i.e. the quadrupole
approximation \cite{landau}), is that the power ${\displaystyle
\frac{dE}{d\Omega}}$, radiated in a solid angle $\Omega$ with
polarization $e_{ij}$, is

\begin{equation}
\frac{dE}{d\Omega}=\frac{G}{8\pi c^{5}}\left(\frac{d^{3}Q_{ij}}{dt^{3}}e_{ij}\right)^{2}\label{eq:P}\end{equation}
where $Q_{ij}$ is the  quadrupole mass tensor
\begin{equation}
Q_{ij}=\sum _a
m_a(3x_a^ix_a^j-\delta_{ij}r_a^2)~,\label{qmasstensor}
\end{equation}
$r_a$  being the modulus of the vector radius of the $a-th$
particle and the sum running over all masses $m_{a}$ in the
system. We must note that the result is independent of the kind of
stresses present into  dynamics. If one sums Eq.(\ref{eq:P}) over
the two allowed polarizations, one obtains
\begin{eqnarray}
\sum_{pol}\frac{dE}{d\Omega} & = & \frac{G}{8\pi c^{5}}\left[\frac{d^{3}Q_{ij}}{dt^{3}}\frac{d^{3}Q_{ij}}{dt^{3}}-2n_{i}\frac{d^{3}Q_{ij}}{dt^{3}}n_{k}\frac{d^{3}Q_{kj}}{dt^{3}}-\frac{1}{2}\left(\frac{d^{3}Q_{ii}}{dt^{3}}\right)^{2}\right.\nonumber \\
 &  & \left.+\frac{1}{2}\left(n_{i}n_{j}\frac{d^{3}Q_{ij}}{dt^{3}}\right)^{2}+
 \frac{d^{3}Q_{ii}}{dt^{3}}n_{j}n_{k}\frac{d^{3}Q_{jk}}{dt^{3}}\right]\,,\label{eq:sommatoria}\end{eqnarray}
where $\hat{n}$ is the unit vector in the  radiation direction.
The total radiation rate is obtained by integrating
Eq.(\ref{eq:sommatoria}) over all emission directions; the result
is

\begin{equation}
P=\frac{dE}{d\Omega}=\frac{G}{c^{5}}\left(\frac{d^{3}Q_{ij}}{dt^{3}}\frac{d^{3}Q_{ij}}{dt^{3}}-
\frac{1}{3}\frac{d^{3}Q_{ii}}{dt^{3}}\frac{d^{3}Q_{jj}}{dt^{3}}\right)\,.\label{eq:potenza}\end{equation}

It is then possible to estimate the amount of energy emitted in
the form of GWs from a system of massive interacting objects. In
this case, the components of the quadrupole mass tensor are
\begin{equation}
\begin{array}{lll}
Q_{xx}=\mu r^2(3\cos{^2\phi}\sin{^2\theta}-1)~,\\ \\
Q_{yy}=\mu r^2(3\sin{^2\phi}\sin{^2\theta}-1)~,\\ \\
Q_{zz}=\frac{1}{2} r^2 \mu  (3 \cos2 \theta+1) ~,\\ \\
Q_{xz}=Q_{zx}=r^2 \mu  (\frac{3}{2} \cos\phi \sin2\theta)~,\\ \\
Q_{yz}=Q_{zy}=r^2 \mu  (\frac{3}{2} \sin 2\theta \sin \phi)~,\\ \\
Q_{xy}=Q_{yx}=r^2 \mu  \left(\frac{3}{2} \sin ^2\theta \sin2\phi\right)~,
\end{array}\label{eq:quadrupoli}
\end{equation}
where the masses $m_{i}$  have  polar coordinates
$\{r_{i}\sin\theta\cos\phi,\; r_{i}\sin\theta\sin\phi,\:
r_{i}\cos\theta\}$ and $\mu$ is the reduced mass.  The origin of
the motions is taken at the center of mass. Such components can be
differentiated with respect to time as in Eq.(\ref{eq:potenza}).
The gravitomagnetic corrections affect, essentially, these
quantities and, consequently, the GW amplitude $h$ and the
radiation rate $P$ as we will see below.

\section{Gravitational wave amplitude with gravitomagnetic corrections}

Direct signatures of gravitational radiation are given by
GW-amplitudes and  waveforms. In other words, the identification
of a GW signal is strictly related to the accurate selection of
the waveform shape by interferometers or any possible detection
tool. Such an achievement could give information on the nature of
the GW source, on the propagating medium, and, in principle, on
the gravitational theory producing such a radiation \cite{Dela}.

Considering the formulas of previous Section,  the GW-amplitude
can be evaluated by
\begin{equation}
h^{jk}(t,R)=\frac{2G}{Rc^4}\ddot{Q}^{jk}~, \label{ampli1}
\end{equation}
$R$ being the distance between the source and the observer and,
due to the above polarizations, $\{j,k\}=1,2$.

From Eq.(\ref{ampli1}), it is straightforward to show that, for a
binary system where $m\ll M$ and orbits have gravitomagnetic
corrections, the Cartesian components of GW-amplitude are
\begin{eqnarray}
\nonumber
h^{xx}=2 \mu  \left[\left(3 \cos ^2\phi \sin ^2\theta-1\right)
   \dot{r}^2+6 r \left( \dot{\theta}\cos ^2\phi \sin 2 \theta
   -\dot{ \phi}\sin ^2\theta \sin2 \phi\right) \dot{r}+\right. \\
\nonumber
   +r\left(\left(3 \cos ^2\phi \sin ^2\theta-1\right)\ddot{r}+3
   r \left(\dot{\theta}^2\cos2 \theta \cos ^2\phi -\dot{\phi} \dot{\theta} \sin2
   \theta \sin2\phi \right. \right.\\
 \left.  \left. \left. -\sin\theta
   \left(\sin\theta \left(\dot{ \phi}^2\cos 2 \phi+\ddot{ \phi} \cos\phi \sin\phi\right)-\ddot{ \theta}\cos\theta \cos ^2\phi
\right)\right)\right)\right]
      \end{eqnarray}

\begin{align}
\nonumber
h^{yy}=&2 \mu \left[\left(3 \sin ^2\theta \sin ^2\phi-1\right)
    \dot{r}^2+6 r \left( \dot{ \phi}\sin 2 \phi  \sin ^2\theta
   + \dot{\theta}\sin 2 \theta  \sin ^2\phi \right)
   \dot{r}\right.+\\
   \nonumber
  &+ r \left(\left(3 \sin ^2\theta  \sin ^2\phi
   -1\right) \ddot{r}+3 r \left( \dot{\theta}^2\cos 2 \theta  \sin ^2\phi
   +\dot{ \phi}
   \dot{\theta}\sin 2 \theta \sin 2 \phi + \right. \right.\\
  +&\left.\left.\left. \sin \theta \left(\ddot{\theta}\cos \theta
   \sin ^2\phi+\sin \theta  \left( \dot{ \phi}^2\cos 2 \phi +\ddot\phi\cos \phi \sin \phi  \right)\right)\right)\right)\right]
\end{align}

\begin{align}
\nonumber
h^{xy}=h^{yx}=&3 \mu  \left[\cos 2 \phi \sin \theta \left(4\dot{\theta} \dot{ \phi} \cos \theta
    + \ddot{ \phi}\sin \theta \right)
   r^2+2 \dot{r} \left(2 \dot{ \phi} \cos 2 \phi  \sin ^2\theta
   +\dot{\theta}\sin 2 \theta \sin 2 \phi  \right)r + \right.\\
   +&\left.\frac{1}{2} \sin 2 \phi  \left(2 \ddot{r} \sin ^2\theta
   +r(t) \left( 2\dot{\theta}^2 \cos 2 \theta-4  \dot{ \phi}^2\sin ^2\theta
   +\ddot{\theta}\sin 2 \theta \right)\right)
   r+ \dot{r}^2\sin ^(\theta \sin 2 \phi \right]
\end{align}
where we are assuming geometrized units. The above formulas have
been obtained from Eqs.(\ref{ddr}), (\ref{ddphi}), (\ref{ddtheta})
The  gravitomagnetic corrections give rise to signatures on the
GW-amplitudes that, in the standard Newtonian orbital motion, are
not present (see for example \cite{DeLaurentis,nucita}). On the
other hand, as discussed in the Introduction, such corrections
cannot be discarded in peculiar situations as dense stellar
clusters or in the vicinity of galaxy central regions.

Finally, the expected strain amplitude turns out to be
$h\simeq(h_{xx}^2+h_{yy}^2+2h_{xy}^2)^{1/2}$. In particular,
considering a monochromatic GW, we  have two independent degrees
of freedom which, in the TT gauge, are $h_{+} = h_{xx} + h_{yy}$
and $h_{\times} = h_{xy} + h_{yx}$. We are going to evaluate these
quantities and results are shown in Figs. \ref{Fig:03},
\ref{Fig:04}, \ref{Fig:05}, \ref{Fig:07}.

\subsection{Numerical results}
\begin{figure}[!ht]
\begin{tabular}{|c|}
\hline
\tabularnewline
\includegraphics[scale=0.5]{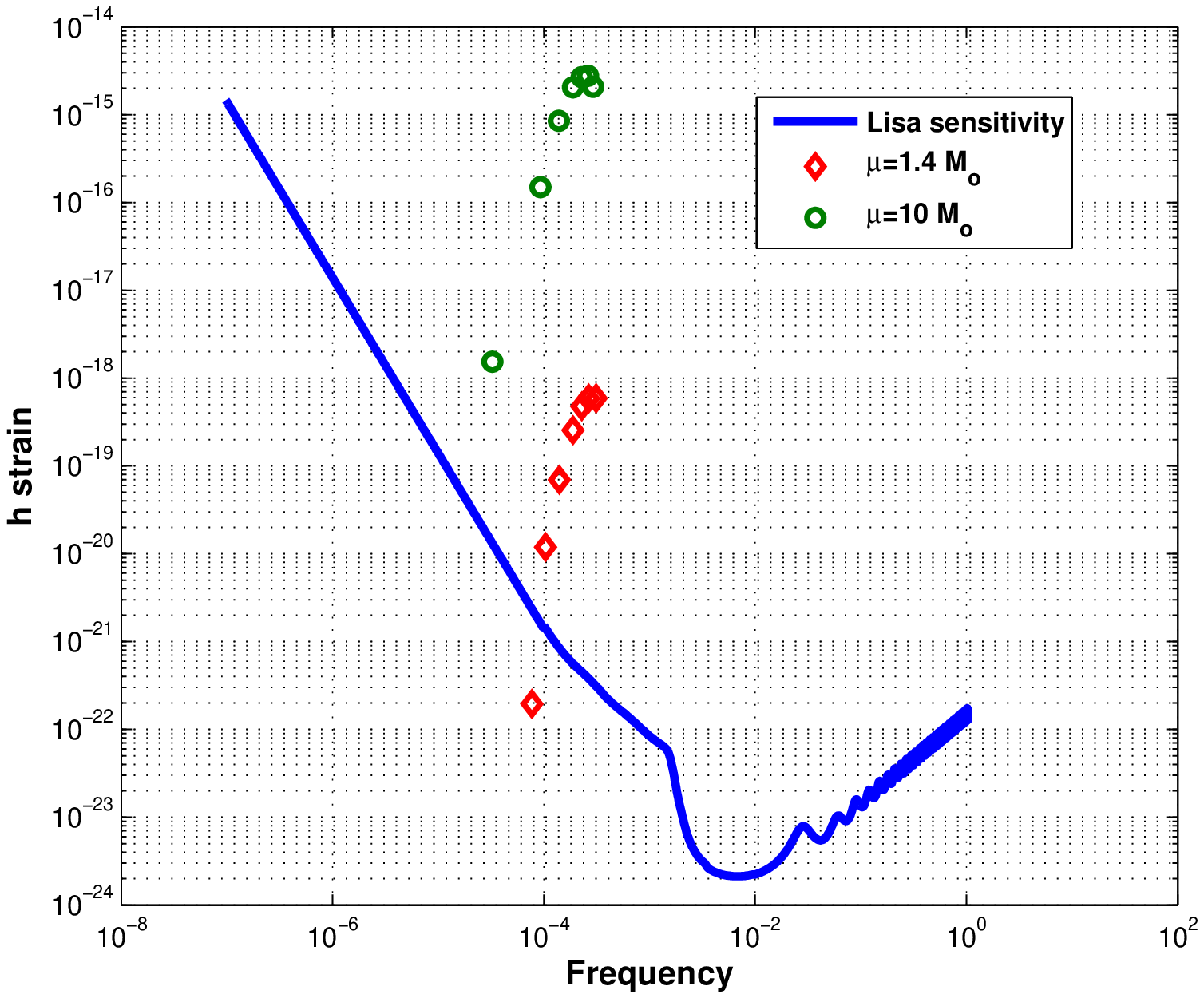}
 \tabularnewline
\hline
\end{tabular}
\caption {Plot of estimated  mean values of GW-emission in terms
of strain $h$ for two binary sources at the Galactic Center
SgrA$^*$ with reduced mass  $\mu\approx1.4M_{\odot}$ (red
diamonds) and $\mu\approx 10M_{\odot}$(green circles). The blue
line is the foreseen LISA sensitivity curve. The waveforms have
been computed for the Earth-distance to SgrA$^*$. The examples we
are showing have been obtained solving the systems for the
parameters and initial conditions reported in  Figs. \ref{Fig:05},
\ref{Fig:07}  and in Table I.}\label{Fig:09}
\end{figure}

Now we have all the ingredients to estimate the effects of
gravitomagnetic corrections on the GW-radiation. Calculations have
been performed in geometrized units in order to evaluate better
the relative corrections in absence of gravitomagnetism. For the
numerical simulations, we have assumed  the fiducial systems
constituted by a $m=1.4M_{\odot}$ neutron star or $m=10M_{\odot}$
massive stellar object orbiting around a MBH $M\simeq 3\times
10^6M_{\odot}$ as  SgrA$^*$. In the extreme mass-ratio limit, this
means that we can consider ${\displaystyle \mu=\frac{mM}{m+M}}$ of
about $\mu \approx 1.4M_{\odot}$  and $\mu \approx10M_{\odot}$.
The computations have been performed  starting with orbital radii
measured in the  mass unit and scaling the distance according to
the values shown in Table I. As it is possible to see in Table I,
starting from $r_{0}=20\mu$ up to $2500\mu$,   the orbital
eccentricity ${\displaystyle
\bm{e}=\frac{r_{max}-r_{min}}{r_{max}+r_{min}}}$ evolves towards a
circular orbit. In  Table I,  the GW-frequencies, in $mHz$, as
well as the $h$ amplitude strains and the two polarizations
$h_{+}$ and $h_{\times}$ are shown. The values are the mean values
of the GW $h$ amplitude strains ($h=\frac{h_{max}+h_{min}}{2}$)
and the maxima of the polarization waves (see Figs. \ref{Fig:05}
and \ref{Fig:07}). In Fig. \ref{Fig:09},   the fiducial LISA
sensitivity curve is shown \cite{LISA} considering the confusion
noise produced by White Dwarf binaries (blue curve). We show also
the $h$ amplitudes (red diamond and green circles for $\mu\approx
1.4 M_{\odot}$ and $\approx 10 M_{\odot}$ respectively). It is
worth noticing that, due to very high Signal to Noise Ratio, the
binary systems which we are considering  result extremely
interesting, in terms of probability detection, for the LISA
interferometer (see Fig. \ref{Fig:09}).

\begin{table}[!ht]
\caption{GW-amplitudes and frequencies as function of eccentricity
$e$, reduced mass $\mu$, orbital radius $r_0$ for the  two cases
of fiducial stellar objects  $m\simeq 1.4 M_{\odot}$ and $m\simeq
10 M_{\odot}$ orbiting around a MBH of mass $M\simeq 3\times
10^6M_{\odot}$.}
\begin{tabular}{|c|c|} \hline
\textbf{$1.4 M_{\odot} $}  & \textbf{$10 M_{\odot} $}\\
\begin{tabular}{|c|c|c|c|c|c}
  \hline
    $\frac{r_{0}}{\mu} $ & $  e $ &  $   f(mHz) $ & $ h $ & $  h_{+} $ & $  h_{\times} $\\
   \hline
  \hline
      $20  $ & $   0.91 $ &  $   7.7\cdot 10^{-2} $ & $ 2.0\cdot 10^{-22} $ & $  5.1\cdot 10^{-23} $ & $  5.1\cdot 10^{-22} $\\
         $200  $ & $  0.79 $ &  $1.1\cdot 10^{-1} $ & $ 1.2\cdot 10^{-20} $ & $  2.2\cdot 10^{-21} $ & $  3.1\cdot 10^{-20} $\\
         $500  $ & $  0.64 $ & $1.4\cdot 10^{-1}$ & $  6.9\cdot 10^{-20}$ & $   8.7\cdot 10^{-21}$ & $   1.7\cdot 10^{-19}$\\
        $1000  $ & $ 0.44 $ & $  1.9\cdot 10^{-1} $ & $ 2.6\cdot 10^{-19} $ & $  6.4\cdot 10^{-20}  $ & $ 6.4\cdot 10^{-19} $\\
        $1500  $ & $ 0.28 $ & $  2.3\cdot 10^{-1} $ & $ 4.8\cdot 10^{-19} $ & $  3.6\cdot 10^{-20} $ & $  1.2\cdot 10^{-18} $\\
        $2000  $ & $ 0.14 $ & $  2.7\cdot 10^{-1} $ & $ 5.9\cdot 10^{-19} $ & $   4.9\cdot 10^{-20} $ & $  1.3\cdot 10^{-18} $\\
        $2500   $ & $ 0.01 $ & $   3.1\cdot 10^{-1} $ & $ 5.9\cdot 10^{-19} $ & $  1.7\cdot 10^{-20} $ & $  9.2\cdot 10^{-19} $\\
\end{tabular}
&
\begin{tabular}{c|c|c|c|c|}
  \hline
   $ e $ &  $   f(mHz) $ & $ h $ & $  h_{+} $ & $  h_{\times} $\\
   \hline
  \hline
      $0.98 $ &  $   3.2\cdot 10^{-2} $ & $ 1.5\cdot 10^{-18} $ & $  1.6\cdot 10^{-19} $ & $  4.3\cdot 10^{-18} $\\
      $0.87 $ & $   9.2\cdot 10^{-2} $ & $ 1.5\cdot 10^{-16} $ & $  2.5\cdot 10^{-18} $ & $  4.1\cdot 10^{-16} $\\
      $0.71 $ & $  1.4\cdot 10^{-1}$ & $  8.5\cdot 10^{-16}$ & $   7.0\cdot 10^{-18}$ & $   2.4\cdot 10^{-15}$\\
      $ 0.49 $ & $  1.9\cdot 10^{-1} $ & $ 2.0\cdot 10^{-15} $ & $  1.6\cdot 10^{-17}  $ & $ 5.6\cdot 10^{-15} $\\
      $ 0.32 $ & $  2.3\cdot 10^{-1} $ & $ 2.7\cdot 10^{-15} $ & $   2.5\cdot 10^{-17} $ & $  7.4\cdot 10^{-15} $\\
      $ 0.19 $ & $  2.6\cdot 10^{-1} $ & $ 2.8\cdot 10^{-15} $ & $   3.3\cdot 10^{-17} $ & $  7.6\cdot 10^{-15} $\\
       $ 0.08 $ & $   2.9\cdot 10^{-1} $ & $ 2.1\cdot 10^{-15} $ & $  4.0\cdot 10^{-17} $ & $  5.6\cdot 10^{-15} $\\

  \end{tabular}
  \\
 \hline
\end{tabular}
\end{table}

\section{Event number estimations towards Sagittarius A$^*$}
At this point, it is important to give some estimates of the
number of events where gravitomagnetic effects could be a
signature for orbital motion and  gravitational radiation.  From
the GW emission point of view, close orbital encounters,
collisions and tidal interactions have to be dealt on average if
we want to investigate the gravitational radiation in a dense
stellar system. On the other hand, dense stellar regions are the
favored target for LISA interferometer \cite{freitag} so it is
extremely useful to provide suitable numbers before its launching.

To this end, it is worth giving the  stellar encounter rate
producing GWs in  astrophysical systems like dense globular
clusters or  the Galactic Center. In general, stars are
approximated as point masses. However, in dense regions of stellar
systems,  a star can pass so close to another that they raise
tidal forces which dissipate their relative orbital kinetic energy
and the Newtonian mechanics or the weak field limit of GR cannot
be adopted as  good approximations. In some cases, the loss of
energy can be so large that stars form binary (the situation which
we have considered here)  or multiple systems; in other cases, the
stars collide and coalesce into a single star; finally stars can
exchange gravitational interaction in non-returning encounters.

To investigate and parameterize all these effects, one has to
compute the collision time $t_{coll}$, where $1/t_{coll}$ is the
collision rate, that is, the average number of physical collisions
that a given star suffers per unit time. As a rough approximation,
one can restrict to stellar clusters in which all stars have the
same mass $m$.

Let us consider an encounter with initial relative velocity
$\mathbf{v}_{0}$ and impact parameter $b$. The angular momentum
per unit mass of the reduced particle is $L=bv_{0}$. At the
distance of closest approach, which we denote by $r_{coll}$, the
radial velocity must be zero, and hence the angular momentum is
$L=r_{coll}v_{max}$, where $v_{max}$ is the relative speed at
$r_{coll}$. It is easy to show that \cite{binney}

\begin{equation}
b^{2}=r_{coll}^{2}+\frac{4Gmr_{coll}}{v_{0}^{2}}\,.\label{eq:b}\end{equation}
If we set $r_{coll}$ equal to the sum of the radii of the two
stars,  a collision will occur if  the impact parameter is less
than the value of $b$, as determined by Eq.(\ref{eq:b}).

The function $f(\mathbf{v}_{a})d^{3}\mathbf{v}_{a}$ gives the
number of stars per unit volume with velocities in the range
$\mathbf{v}_{a}+d^{3}\mathbf{v}_{a}.$ The number of encounters per
unit time with impact parameter less than $b$, which are suffered
by a given star, is  $f(\mathbf{v}_{a})d^{3}\mathbf{v}_{a}$ times
the volume of the annulus with radius $b$ and length $v_{0}$, that
is,

\begin{equation}
\int f(\mathbf{v}_{a})\pi
b^{2}v_{0}d^{3}\mathbf{v}_{a}\label{eq:integrale}\end{equation}
where $v_{0}=\left|\mathbf{v-v}_{a}\right|$ and $\mathbf{v}$ is
the velocity of the considered star. The quantity in
Eq.(\ref{eq:integrale}) is equal to $1/t_{coll}$ for a star with
velocity $\mathbf{v}$: to obtain the mean value of $1/t_{coll}$,
we average over $\mathbf{v}$ by multiplying (\ref{eq:integrale})
by $f(\mathbf{v})/\nu$, where $\nu=\int
f(\mathbf{v})d^{3}\mathbf{v}$ is the number density of stars and
the integration is over $d^{3}\mathbf{v}$. Thus
{}``\begin{equation}
\frac{1}{t_{coll}}=\frac{\nu}{8\pi^{2}\sigma^{6}}\int
e^{-(v^{2}+v_{a}^{2})/2\sigma^{2}}\left(r_{coll}\left|\mathbf{v-v}_{a}\right|+
\frac{4Gmr_{coll}}{\left|\mathbf{v-v}_{a}\right|}\right)d^{3}\mathbf{v}d^{3}\mathbf{v}_{a}\,.
\label{eq:invtcoll}\end{equation} Replacing the variable
$\mathbf{v}_{a}$ by $\mathbf{V}=\mathbf{v-v}_{a}$, the argument of
the exponential is then
$-\left[\left(\mathbf{v}-\frac{1}{2}\mathbf{V}\right)^{2}+\frac{1}{4}V^{2}\right]/\sigma^{2}$,
and if we replace the variable $\mathbf{v}$ by ${\displaystyle
\mathbf{v}_{cm}=\mathbf{v}-\frac{1}{2}\mathbf{V}}$ (the center of
mass velocity), then one has

\begin{equation}
\frac{1}{t_{coll}}=\frac{\nu}{8\pi^{2}\sigma^{6}} \int
e^{-(v_{cm}^{2}+V^{2})/2\sigma^{2}}\left(r_{coll}V+
\frac{4Gmr_{coll}}{V}\right)dV\,.\label{eq:invtcoll1}\end{equation}
The integral over $\mathbf{v}_{cm}$ is given by

\begin{equation}
\int
e^{-v_{cm}^{2}/\sigma^{2}}d^{3}\mathbf{v}_{cm}=\pi^{3/2}\sigma^{3}\,.\label{eq:intint}\end{equation}
Thus

\begin{equation}
\frac{1}{t_{coll}}=\frac{\pi^{1/2}\nu}{2\sigma^{3}}\int_{\infty}^{0}e^{-V^{2}/4\sigma^{2}}
\left(r_{coll}^{2}V^{3}+4GmVr_{coll}\right)dV\label{eq:invtcoll2}\end{equation}
The integrals can be  easily calculated and then we find

\begin{equation}
\frac{1}{t_{coll}}=4\sqrt{\pi}\nu\sigma
r_{coll}^{2}+\frac{4\sqrt{\pi}\nu
Gmr_{coll}}{\sigma}\,.\label{eq:invtcooll3}\end{equation} The
first term of this result can be derived from the kinetic theory.
The rate of interaction is $\nu\Sigma\left\langle V\right\rangle$,
where $\Sigma$ is the cross-section and $\left\langle
V\right\rangle $ is the mean relative speed. Substituting
$\Sigma=\pi r_{coll}^{2}$ and $\left\langle V\right\rangle
=4\sigma/\sqrt{\pi}$ (which is appropriate for a Maxwellian
distribution whit dispersion $\sigma$) we recover the first term
of (\ref{eq:invtcooll3}). The second term represents the
enhancement in the collision rate by gravitational focusing, that
is, the deflection of trajectories by the gravitational attraction
of the two stars.

If $r_{*}$ is the stellar radius, we may set $r_{coll}=2r_{*}$. It
is convenient to introduce the escape speed from stellar surface,
${\displaystyle v_{*}=\sqrt{\frac{2Gm}{r_{*}}}}$, and to rewrite
Eq.(\ref{eq:invtcooll3}) as

\begin{equation}
\Gamma=\frac{1}{t_{coll}}=16\sqrt{\pi}\nu\sigma
r_{*}^{2}\left(1+\frac{v_{*}^{2}}{4\sigma^{2}}\right)=16\sqrt{\pi}\nu\sigma
r_{*}^{2}(1+\Theta),\label{eq:invtcoll4}\end{equation}
where

\begin{equation}
\Theta=\frac{v_{*}^{2}}{4\sigma^{2}}=\frac{Gm}{2\sigma^{2}r_{*}}\label{eq:safronov}\end{equation}
is the Safronov number \cite{binney}. In evaluating the rate, we
are considering only those  encounters producing gravitational
waves, for example,  in the LISA range, i.e. between $10^{-4}$ and
$10^{-1}$ Hz (see e.g. \cite{Rub}). Numerically, we have
\begin{equation}
\Gamma \simeq  5.5\times 10^{-10} \left(\frac{v}{10 {\rm km
s^{-1}}}\right) \left(\frac{\sigma}{UA^2}\right) \left(\frac{{\rm
10 pc}}{R}\right)^3 {\rm
yrs^{-1}}\qquad\Theta<<1\label{eq:thetamin}
\end{equation}
\begin{equation}
\Gamma \simeq  5.5\times 10^{-10} \left(\frac{M}{10^5 {\rm
M_{\odot}}}\right)^2 \left(\frac{v}{10 {\rm km s^{-1}}}\right)
\left(\frac{\sigma}{UA^2}\right) \left(\frac{{\rm 10
pc}}{R}\right)^3 {\rm yrs^{-1}}\qquad\Theta>>1\label{eq:thetamagg}
\end{equation}
If $\Theta>>1$, the energy dissipated exceeds the relative kinetic
energy of the colliding stars, and the stars  coalesce into a
single star. This new star may, in turn, collide and merge with
other stars, thereby becoming very massive. As its mass increases,
the collision time is shorten and then there may be runaway
coalescence leading to the formation of a few supermassive objects
per clusters. If $\Theta<<1$, much of the mass in the colliding
stars may be liberated and forming new stars or a single
supermassive objects (see \cite{Belgeman,Shapiro}). Both cases are
interesting for LISA purposes.

Note that when one has the effects of quasi-collisions (where
gravitomagnetic effects, in principle, cannot be discarded) in an
encounter of two stars in which the minimal separation is several
stellar radii, violent tides will raise on the surface of each
star. The energy that excites the tides comes from the relative
kinetic energy of the stars. This effect is important for
$\Theta>>1$ since the loss of small amount of kinetic energy may
leave the two stars with negative total energy, that is, as a
bounded  binary system. Successive peri-center passages will
dissipates more energy by GW radiation, until the binary orbit is
nearly circular with a negligible or null GW radiation emission.

Let us apply these considerations to the Galactic Center which can
be modelled as a system of several compact stellar clusters, some
of them similar to very compact globular clusters with high
emission in X-rays \cite{townes}.

For a typical globular cluster  around the Galactic Center, the
expected event rate is of the order of $2\times 10^{-9}$
yrs$^{-1}$ which may be increased at least by a factor $\simeq
100$ if one considers the number of globular clusters in the whole
Galaxy. If the stellar cluster at the Galactic Center is taken
into account and assuming the total mass $M\simeq 3\times 10^6$
M$_{\odot}$, the velocity dispersion $\sigma\simeq $ 150 km
s$^{-1}$ and the radius of the object $R\simeq$ 10 pc (where
$\Theta=4.3$), one expects to have $\simeq 10^{-5}$ open orbit
encounters per year. On the other hand, if a cluster with total
mass $M\simeq 10^6$ M$_{\odot}$, $\sigma\simeq $ 150 km s$^{-1}$
and $R\simeq$ 0.1 pc is considered, an event rate number of the
order of unity per year is obtained. These values could be
realistically achieved by data coming from the forthcoming space
interferometer LISA. As a secondary effect, the above  wave-forms
could constitute the "signature"  to classify the different
stellar encounters  thanks to the differences of the shapes (see
Figs. \ref{Fig:05} and \ref{Fig:07}).

\section{Conclusions}
In this paper we have discussed  gravitomagnetic effects on
orbital that could give rise to interesting phenomena in tight
binding systems such as binaries of evolved objects (neutron stars
or black holes). The effects become particularly relevant when
such objects orbit around or fall toward very MBHs as those at the
center of galaxies. The effects reveal particularly interesting if
$v/c$ is in the range $(10^{-1}\div 10^{-4})c$.  Gravitomagnetic
orbital corrections, after long integration time, induce
precession and nutation effects capable of affecting the stability
basin of the orbits. The global structure of such a basin is
extremely sensitive to the initial radial velocities and angular
velocities, the initial energy and masses which can determine
possible transitions to chaotic behavior. In principle, GW
emission could present signatures of gravitomagnetic corrections
after suitable integration times in particular for the on-going
LISA space laser interferometric GW antenna.


\begin{thebibliography}{99}

\bibitem{Abra} A. Abramovici et al., {\it Science} {\bf 256} (1992) 325; http://www.ligo.org

\bibitem{Caron} B. Caron et al., {\it Class. Quant. Grav.} {\bf 14} (1997) 1461; http://www.virgo.infn.it

\bibitem{Luck} H. Luck et al., {\it Class. Quant. Grav.} {\bf 14} (1997) 1471; http://www.geo600.uni-hannover.de

\bibitem{Ando} M. Ando et al., {\it Phys. Rev. Lett.} {\bf 86} (2001) 3950; http://tamago.mtk.nao.ac.jp

\bibitem{LISA} http://www.lisa-science.org

\bibitem{Poisson}  E. Poisson, {\it Living Rev. Rel.} {\bf 6} (2004) 3, and references therein.
http://relativity.livingreviews.org

\bibitem{Mino} Y.Mino, {\it Prog. Theor. Phys.} {\bf 113} (2005) 733; ibid. {\it Prog. Theor. Phys.} {\bf 115} (2006) 43;
 A. Pound, E. Poisson, and B.G. Nickel, {\it Phys. Rev.} {\bf D 72} (2005) 124001; L. Barack and C. Lousto, {\it Phys.
Rev.} {\bf D 72} (2005) 104026; L. Barack and N. Sago, {\it Phys.
Rev.} {\bf D 75} (2007) 064021.

\bibitem{Finn} L. S. Finn and K. S. Thorne, {\it Phys. Rev.} {\bf D 62} (2000) 124021;
S. Babak et al. {\it Phys. Rev.} {\bf D 75} (2007) 024005; G.
Sigl, J. Schnittman, and A. Buonanno, {\it Phys. Rev.} {\bf D 75}
(2007) 024034.

\bibitem{viollier}
R.D. Viollier {\it Prog. Part. Nucl. Phys.} {\bf 32} (1994) 51.

\bibitem{pla} S. Capozziello, G. Iovane, {\it Phys. Lett.} {\bf A 259}
 (1999) 185.

\bibitem{Sigurdsson} S. Sigurdsson and M. Rees, {\it Mont. Not. R. Astron. Soc.}, {\bf 284} (1997) 318.

\bibitem{Sigurdsson2} S. Sigurdsson, {\it Class. Quant. Grav.} {\bf 14} (1997) 1425.

\bibitem{Danzmann} K. Danzmann et al., LISA- Laser Interferometer Space Antenna, Pre-Phase A Report,
Max-Planck-Institut fur Quantenoptic,Report MPQ 233 (1998).

\bibitem{freitag}
P. Amaro-Seoane et al., {\it Class. Quant. Grav.} {\bf 24} (2007)
R113.

\bibitem{Capoz}
S. Capozziello, M. De Laurentis, F. Garufi and L. Milano, {\it
Phys. Scr.} {\bf 79} (2009) 025901.

\bibitem{binney}
J.Binney and S.Tremaine, {\it Galactic Dynamics},
 Princeton University Press, Princeton, New Jersey (1987).

\bibitem{landau}
L. Landau and E.M. Lifsits, {\it Mechanics}, Pergamon Press, New
York  (1973).

\bibitem{Thirring1} H. Thirring, {\it Phys. Z.} {\bf 19} (1918) 204

\bibitem{Thirring} H. Thirring, Phys. Z., 19, (1918) 33; 22, (1921) 29; J. Lense and H.
Thirring, Phys. Z., 19, (1918) 156. The english translation can be
found in B. Mashhoon, F.W. Hehl and D.S. Theiss, Gen. Rel. Grav.,
16, (1984) 711.

\bibitem{iorio}
L. Iorio and V. Lainey, {\it Int. Jou. Mod. Phys. D} {\bf 14}
(2005) 2039.

\bibitem{Ryan} F. D. Ryan, Phys. Rev. D 56 (1997) 1845.

\bibitem{DeLaurentis}
S. Capozziello and M. De Laurentis, {\it Astrop. Phys.} {\bf 30}
(2008) 105.

\bibitem{nucita}
S. Capozziello, M. De Laurentis, F. De Paolis, G. Ingrosso and A.
Nucita, {\it Mod. Phys. Lett.} {\bf A 23}(2008) 99.

\bibitem{Cutler1}
L. Barack and C. Cutler, {\it Phys. Rev.} {\bf D 69} (2004)
082005.

\bibitem{Cutler2}
L. Barack and C. Cutler, {\it Phys. Rev.} {\bf D 70} (2004)
122002.

\bibitem{FreitagApJ} M. Freitag, Astro.\ J.\ {\bf 583} (2003) L21.

\bibitem{Dela} S. Capozziello,  M. De Laurentis, M. Francaviglia,   {\it Astrop. Phys.}  {\bf 29} (2008) 125.

\bibitem{Rub} L.J. Rubbo, K. Holley - Bockelmann, and L.S. Finn, {\it Ap.J.} {\bf 649} (2006) L25.

\bibitem{Belgeman}M.C. Belgeman and M.J. Rees \emph{Mon. Not. Roy. Astron.
Soc.,} {\bf 185} (1978) 847.

\bibitem{Shapiro}A.P. Lightman and S. L. Shapiro \emph{Rev. Mod. Phys.,}
{\bf 50} (1978) 437.

\bibitem{townes} R. Genzel and C.H. Townes, {\it Ann. Rev. Astron.
Astroph.} {\bf 25} (1987) 1.

\end{thebibliography}
\end{document}